\documentclass[aps,twocolumn,prx,preprintnumbers,showpacs,amsmath,amssymb,superscriptaddress]{revtex4-1}
\usepackage[pdftex, colorlinks, citecolor=blue]{hyperref}   

\usepackage{graphicx}
\usepackage{dcolumn}
\usepackage{multirow}
\usepackage{booktabs}
\usepackage{txfonts}
\usepackage{xcolor}
\usepackage{bm}
\usepackage{amssymb}
\usepackage{amsmath}
\usepackage{latexsym}
\usepackage{epsfig}
\usepackage{amsbsy}
\usepackage{array}
\usepackage{tabularx}
\usepackage{tabularray}
\usepackage{threeparttable}
\usepackage{esvect} 
\usepackage{extarrows}
\usepackage{float}
\usepackage[british]{babel}
\usepackage{soul}

\renewcommand{\arraystretch}{1.5}
\newcommand{\C}[1]{C_{#1}}
\newcommand{\AleNdt}[0]{\mathrm{Al}_{11}\mathrm{Nd}_{3}}
\newcommand{\AltwNd}[0]{\mathrm{Al}_{2}\mathrm{Nd}}
\newcommand{\AlthrNd}[0]{\mathrm{Al}_{3}\mathrm{Nd}}
\newcommand{\AlNd}[0]{\mathrm{AlNd}}
\newcommand{\rd}[1]{\textcolor{red}{#1}}
\newcommand{\Al}[1]{Al\textsubscript{#1}}
\newcommand{\Nd}[1]{Nd\textsubscript{#1}}

\begin{document}
\title{Deformability, inherent mechanical properties and chemical bonding of $\mathbf{Al_{11}Nd_3}$ in Al-Nd target material}

\author{Xue-Qian Wang}
\affiliation{%
Key Laboratory for Anisotropy and Texture of Materials (Ministry of Education), School of Material Science and Engineering, Northeastern University, Shenyang 110819, China.
}%
\author{Run-Xin Song}
\affiliation{%
Key Laboratory for Anisotropy and Texture of Materials (Ministry of Education), School of Material Science and Engineering, Northeastern University, Shenyang 110819, China.
}%

\author{Xu Guan}
\affiliation{%
Key Laboratory for Anisotropy and Texture of Materials (Ministry of Education), School of Material Science and Engineering, Northeastern University, Shenyang 110819, China.
}%

 \author{Shuan Li}
 \affiliation{%
 National Engineering Research Center for Rare Earth, GRIREM Advanced Materials Co., Ltd., Beijing 100088, China
 }%

\author{Shuchen Sun}
\affiliation{%
School of Metallurgy, Northeastern University, Shenyang 110819, China.
}%

 \author{Hongbo Yang}
 \affiliation{%
 National Engineering Research Center for Rare Earth, GRIREM Advanced Materials Co., Ltd., Beijing 100088, China
 }%

 \author{Daogao Wu}
 \email{wudaogao@grirem.com (D.G. Wu)}
 \affiliation{%
 National Engineering Research Center for Rare Earth, GRIREM Advanced Materials Co., Ltd., Beijing 100088, China
 }%

\author{Ganfeng Tu}
\affiliation{%
School of Metallurgy, Northeastern University, Shenyang 110819, China.
}%

\author{Song Li}
\email{lis@atm.neu.edu.cn (S. Li)}
\affiliation{%
Key Laboratory for Anisotropy and Texture of Materials (Ministry of Education), School of Material Science and Engineering, Northeastern University, Shenyang 110819, China.
}%

\author{Hai-Le Yan}
\email{yanhaile@mail.neu.edu.cn (H.-L. Yan)}
\affiliation{%
Key Laboratory for Anisotropy and Texture of Materials (Ministry of Education), School of Material Science and Engineering, Northeastern University, Shenyang 110819, China.
}%

\author{Liang Zuo}
\affiliation{%
Key Laboratory for Anisotropy and Texture of Materials (Ministry of Education), School of Material Science and Engineering, Northeastern University, Shenyang 110819, China.
}%

\begin{abstract}
Microstructure uniformity of the Al-Nd target materials with $\AleNdt$ significantly affects the performance of the fabricated film, which is widely used as wiring material in large-size thin-film transistor liquid crystal display (TFT-LCD) panels. Understanding the inherent mechanical properties and chemical bonds of $\AleNdt$ is crucial for homogenizing the Al-Nd target. Here, by a combined experimental and \textit{ab-initio} theoretical study, the microstructure and deformability of the Al-3wt\%Nd alloy and the inherent mechanical properties and chemical bonds of $\AleNdt$ are investigated comprehensively. The Al-3wt\%Nd alloy is composed of the pre-eutectic $\alpha$-Al matrix and the eutectic $\alpha$-Al and a high stable $\alpha$-$\AleNdt$ phases. During the plastic deformation, the eutectic microstructure transforms from a cellular to a lamellar shape, while the morphology and dimension of $\alpha$-$\AleNdt$ are not changed significantly. By examining ideal tensile strength, elastic moduli, hardness and brittleness-ductility, the hardness-brittleness of $\alpha$-$\AleNdt$ is quantitatively evaluated, accounting for its difficulties of plastic deformation and fragmentation. Combining band structure, population analysis, topological analysis and crystal orbital Hamilton population, it is revealed that $\alpha$-$\AleNdt$ possesses two types of chemical bonds: the Nd-Al and Al-Al bonds. The former is a typical ionic bond with electron transfer from Nd to Al, while the latter, dominated by both 3\textit{s}-3\textit{p} and 3\textit{p}-3\textit{p} interactions, is a weak covalent bond. The mixed chemical bond is responsible for the high hardness-brittleness of $\alpha$-$\AleNdt$. This work is expected to lay a foundation for Al-Nd alloy and catalyze the fabrication of high-quality Al-Nd target materials.
\\ %
\end{abstract}

\maketitle
\section{Introduction}

Aluminum is widely used as wiring material in large-size, high-response and high-precision thin-film transistor liquid crystal display (TFT-LCD) panels due to its advantages of low electron resistivity and low cost~\cite{RN658, RN655,RN656,RN649,RN657}. Instead of pure Al, the Al-3wt\%Nd alloy has come into practical use since adding a small amount of Nd can effectively suppress the formation of harmful hemispheric hillocks during the heating process~\cite{RN655, Onishi1996InfluenceOA, Onishi19972339}. In industry, the Al-Nd wiring materials in TFT-LCD are fabricated with the magnetron sputtering technique by using the Al-Nd alloy as the target material.  It is reported that the microstructure uniformity of the Al-Nd target significantly affects the overall performance of the fabricated Al-Nd film~\cite{RN654, RN649}. In the Al-Nd target material, the Nd element mainly exists in the form of eutectic $\AleNdt$ phase~\cite{zujun2017}. The challenge of homogenizing the microstructure of Al-Nd target material lies just in this eutectic phase. Therefore, uncovering the phase stability, deformability, inherent mechanical properties and chemical bonds of $\AleNdt$ is inevitably necessary to design and fabricate high-quality Al-Nd target materials and further advanced large-size TFT-LCD panels.

Unlike the extensive studies on other Al-Nd compounds, such as $\AlthrNd$, $\AltwNd$ and $\AlNd$~\cite{RN452,RN619,RN591,RN624}, the knowledge of $\AleNdt$ is extremely limited in the open literature. From the Al-Nd phase diagram~\cite{zhang2010phase}, it is known that $\AleNdt$ has two kinds of crystal structures, \textit{i.e.}, the high-temperature $\beta$ phase and the low-temperature $\alpha$ phase, with a transitional temperature around 950 C\textdegree. S. Lv and coworkers~\cite{RN590} studied the thermodynamic stabilities of a series of Al$_{11}$RE$_3$ where RE represents the rare earth element, demonstrating that the $\AleNdt$ phase exhibits high thermodynamic stability against the decomposition into Al$_2$Nd and Al at temperatures below 1000 K. H. Yamamoto and coworkers~\cite{RN593} determined the standard formation entropy of $\alpha$-$\AleNdt$ by heat capacity measurement from near absolute zero Kelvin.  Very recently, T. Fan and coworkers~\cite{RN595} studied the elastic properties of several Al$_{11}$RE$_3$ (RE= La, Ce, Pr, Nd and Sm) compounds, claiming that Al$_{11}$RE$_3$ can be utilized as the strengthening phase to effectively improve the mechanical properties of high-performance heat-resistant Al alloys.  Nevertheless, until now, there is still no comprehensive knowledge about the stability, deformability, and inherent mechanical properties of the $\AleNdt$ phase in the Al-3wt\%Nd alloy. Furthermore, the electron structure and chemical bonding information of the $\AleNdt$ phase, which is essential for understanding the inherent mechanical properties of materials, has not been elucidated.

To bridge these knowledge gaps, by a combined experimental and \textit{ab-initio} theoretical study, the microstructure and deformability of the Al-3wt\%Nd alloy and the inherent mechanical properties, electron structures and chemical bonds of $\AleNdt$ are investigated systematically. In Section~\ref{sec:mcrstct}, the crystal structure, chemical composition, and morphological and crystallographic orientation features of the microstructure of as-cast Al-3wt\%Nd alloy are investigated. The thermodynamics and elastic stabilities of $\AleNdt$ are then evaluated. In Section~\ref{sec:dfmblt}, the deformability, intrinsic mechanical strength and brittleness-ductility of $\AleNdt$ in the Al-3wt\%Nd alloy are examined. The deformability is evaluated by tracking the microstructure evolution during the asymmetric cold rolling. The intrinsic mechanical strength is examined by ideal tensile strength ($\sigma_\mathrm{max}$), bulk modulus ($B$), Young’s modulus ($E$), shear modulus ($G$) and hardness. The inherent brittleness-ductility is probed by different brittleness-ductility criteria, including  Pugh's ratio (\textit{G}/\textit{B}), Cauchy pressure ($C_\mathrm{p}$), and Poisson's ratio $\nu$. In Section~\ref{sec:elecstru}, the electronic structures and chemical bonds of $\AleNdt$ are investigated systematically. First, the local chemical environment, which decides electron structure and chemical bonding, is analyzed.  Second, the electron band structure (BS) and density of states (DOS) are calculated. Third, the atom charge population analysis, quantum theory of atoms in molecular (QTAIM) and electron localization function (ELF) topological analysis, crystal orbital Hamilton population (COHP)~\cite{RN577, RN34} and crystal orbital bond index (COBI)~\cite{muller2021crystal1} are studied.

\section{Methodology}

\subsection{Experimental details}
An ingot of the Al-3wt\%Nd alloy was fabricated using vacuum induction melting technology. The samples with an initial thickness of 20 mm were used for asymmetrical rolling at a speed ratio of 1.3. The rolling was carried out at room temperature and was cooled in ice water between passes. After rolling, the sheets with thickness reductions of 25\% and 50\% were obtained. For crystal structure and microstructure characterization, the rectangular samples with dimensions of 5×5×1 mm$^3$ were cut out. To remove surface scratches, the samples were mechanically ground with sandpapers and polished with MgO suspension. The electrolytic polishing technique was then utilized to remove the strained layer near the surface. The solution of electrolytic polishing is a mixture of 20\% HClO$_4$ and 80\% C$_2$H$_5$OH. The voltage and time are 20 V and 20 s, respectively. The crystal structure was examined using X-ray diffraction (XRD, Rigaku Smartlab) with Cu-$K\alpha$ radiation.  To examine the microstructure with the optical microscope, the specimen was chemically etched to highlight the grain boundary network. Keller’s reagent with a solution ratio of 2.5\% HNO$_3$, 1.5\% HCl, 1.0\% HF, 95\% H$_2$O in volume was used. The morphology and the crystallographic orientation were measured by scanning electron microscope (SEM, JSM 7001F) with an electron backscatter diffraction (EBSD) acquisition camera.   

\subsection{Calculation methods}

First-principles calculations were performed using density functional theory (DFT) implemented in the Vienna \textit{ab-initio} Simulation Package (VASP 6.3)~\cite{RN459,RN581}. The Perdew-Burke-Ernzerhof (PBE) parametrization of generalized gradient approximation (GGA) was employed to describe the exchange-correlation function. The electron-ion interactions were described by the projector augmented wave (PAW) pseudopotential approach. The valence electron configurations of $3s^23p^1$ for Al and $5s^25p^64f^46s^2$ for Nd were adopted.  A kinetic energy cutoff of 500 eV was adopted for wave-function expansion. The $k$-point meshes with an interval of 2$\pi$×0.02 Å$^{-1}$ for the Brillouin zone to ensure that the total energy converges within $10^{-6}$ eV/atom. During the structural relaxation, the Hellman-Feynman force on each atom was relaxed to be less than $10^{-3}$ eV/Å. 

The independent elastic constants were determined by computing the second-order derivatives of the total energy with respect to the position of the ions using a finite difference approach~\cite{RN582}. The ideal tensile stress-strain curves along the $a$, $b$, and $c$ axes of orthorhombic lattice of $\AleNdt$ were calculated with the method detailed in Ref.~\cite{guo2024}.  For the electronic structure calculation, the tetrahedron method with the Blöchl correction~\cite{RN573} was used to integrate the Brillouin zone and a denser $k$-point mesh with an interval of 2$\pi$×0.01 Å$^{-1}$ was adopted. Crystalline orbital Hamiltonian population (COHP) and crystal orbital
bond index (COBI) were calculated by using the local orbital basis suite towards electronic-structure reconstruction (LOBSTER) program~\cite{RN570,RN569}. The $3s$ and $3p$ for Al and $5s$, $5p$, $4f$ and $6s$ for Nd were taken as basis sets. The atom charge analyses were performed using the Bader, Mulliken and L\"{o}wdin decomposition techniques~\cite{RN604,RN570,RN569}. The local chemical environment was analyzed with robocrystallographer toolkit~\cite{ganose_jain_2019}. The topological analyses of charge density were performed using the CRITIC2 software package~\cite{otero2014critic2}. The formation energy ($E_{\mathrm{f}}$) of $\alpha$-$\AleNdt$ was calculated by  
\begin{equation}\label{eq:ef}
    E_{\mathrm{f}}=\frac{E_{\mathrm{total}}-N_{\mathrm{Al}} \cdot E_{\mathrm{solid}}^{\mathrm{Al}}-N_{\mathrm{Nd}} \cdot E_{\mathrm{solid}}^{\mathrm{Nd}}}{N_{\mathrm{Al}}+N_{\mathrm{Nd}}}
\end{equation}
where $E_{\mathrm{total}}$ is the total energy, $E_{\mathrm{solid}}^{\mathrm{Al}}$ and $E_{\mathrm{solid}}^{\mathrm{Nd}}$ are the averaging energies per Al and Nd atom in their ground-state structures, respectively. Here, the face-centered cubic (fcc) structure with a space group of \textit{F}m$\overline{3}$m for Al and the $\alpha$-La-type hexagonal structure with a space group of \textit{P}6$_3$/mmc for pure Nd were adopted. $N_{\mathrm{Al}}$ and $N_{\mathrm{Nd}}$ are the atom numbers of Al and Nd at the adopted structural models, respectively. 

\section{Results}

\subsection{Microstructure and phase stability}\label{sec:mcrstct}
\subsubsection{Microstructure and crystal structure}

\begin{figure*}
    \centering
    \includegraphics[width=1\linewidth]{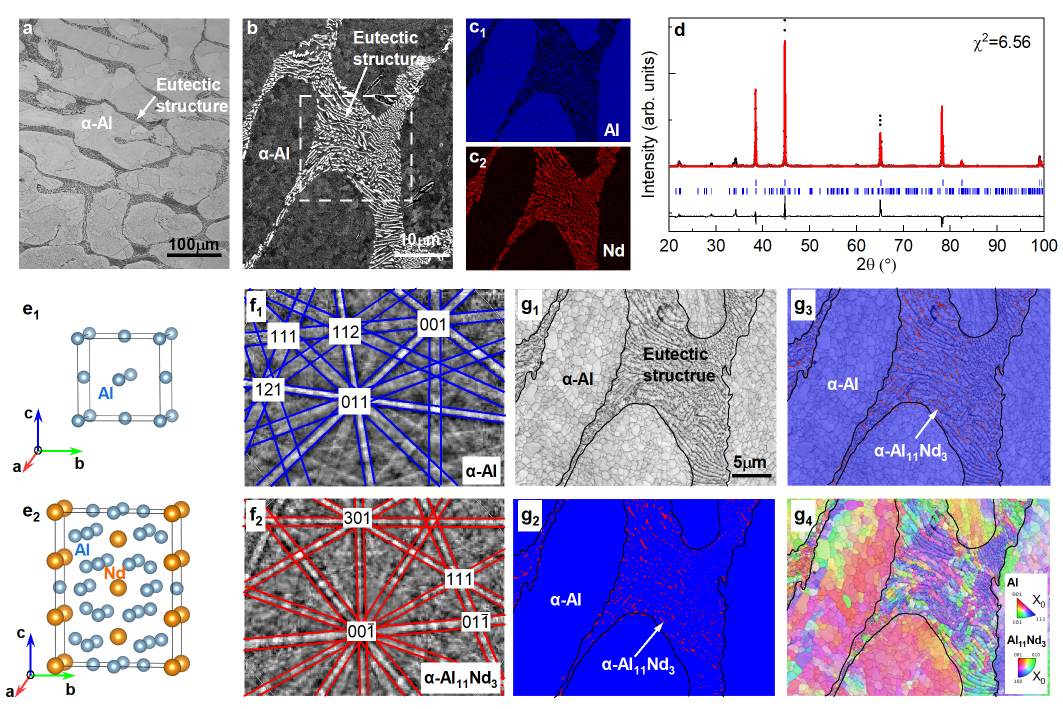}
    \caption{\textbf{Microstructure of the as-casted Al-3wt\%Nd alloy.} (\rd{a}) Optical image. (\rd{b}) Backscattered electron (BSE) image. (\rd{c$_1$}-\rd{c$_2$}) Distributions of Al and Nd for the region indicated in the dashed box in \rd{b}. (\rd{d}) Rietveld refinement of XRD pattern of the bulk sample. (\rd{e$_1$}-\rd{e$_2$}) Crystal structures of $\alpha$-Al and $\alpha$-$\AleNdt$. (\rd{f$_1$}-\rd{f$_2$}) Measured and simulated \textit{Kikuchi} patterns for $\alpha$-Al and $\alpha$-$\AleNdt$. (\rd{g$_1$}-\rd{g$_4$}) Band contrast (BC), phase distribution (PD), stacked BC and PD, and crystallographic orientation map colored using the index of inverse pole figure along the horizontal direction of the sample (X$_0$). The boundaries between the $\alpha$-Al matrix and the eutectic phases are highlighted in bold black lines. The probing region in \rd{g$_1$}-\rd{g$_4$} is exactly the same as that of \rd{c$_1$}-\rd{c$_2$}. Note that some subgrains ($\sim$1 $\mu$m) are detected in the Al matrix, which might be associated with the prominent thermal stress generated during the rapid cooling process.}
    \label{fig:microstru}
\end{figure*}

Figs.~\ref{fig:microstru} shows the microstructure and crystal structure of the casted Al-3wt\%Nd alloy. The sample is composed of the pre-eutectic Al matrix with a grain size around 100 $\mu$m and the cellular eutectic microstructure with a width of 6$\sim$35 $\mu$m (Figs.~\ref{fig:microstru}\rd{a} and \rd{b}). The volume fraction of the Al matrix and the eutectic microstructure are measured to be around 92\% and 8\%, respectively, which is in good agreement with the theoretical value (90.6 wt.\% for the pre-eutectic Al matrix) deduced from the Al-Nd equilibrium phase diagram~\cite{zhang2010phase}. Composition examination shows that the eutectic microstructure is composed of an enriched Nd phase with a composition of Al$_{11}$Nd$_{3}$ (wt\%, Fig.~\ref{fig:microstru}\rd{c$_2$}) and a pure Al phase (Fig.~\ref{fig:microstru}\rd{c$_1$}). The fraction of Al$_{11}$Nd$_{3}$ is around 50\% in the eutectic microstructure, in line with the deduced theoretical value (46.4\% in weight ratio)~\cite{zhang2010phase}. The Al$_{11}$Nd$_{3}$ phase is in a rod shape with a radius of around 0.5$\sim$4 $\mu$m and a length of around 0.6$\sim$19 $\mu$m (Fig.~\ref{fig:microstru}\rd{b} and Supplementary Materials Fig.~\rd{S1}). Note that in both the Al matrix and the Al phase in the eutectic microstructure, almost no Nd element is detected, which is in good agreement with the limited solid solubility of Nd in Al~\cite{zhang2010phase}. By the Rietveld full-profile refinement, as shown in Fig.~\ref{fig:microstru}\rd{d}, the Al matrix and the Al phase in the eutectic microstructure are identified to possess the face-centered cubic (fcc) structure with a space group of \textit{F}m$\overline{3}$m (group number: 225, Fig.~\ref{fig:microstru}\rd{e$_1$}), \textit{i.e.}, the $\alpha$-Al phase. The lattice parameter is determined to be $a$=4.0496(92) \r{A}. This value is very close to that of pure Al (4.0494 \r{A}~\cite{1994Structure}), aligning well with the limited Nd solid solubility in Al. The $\AlthrNd$ phase is identified to possess an orthorhombic structure with a space group of \textit{I}mmm (group number: 71, Fig.~\ref{fig:microstru}\rd{e$_2$}), \textit{i.e.}, $\alpha$-$\AleNdt$ following the nomenclature of Refs.~\cite{RN618,zujun2017}. The lattice parameters are refined to be \textit{a}= 4.3648(41) \r{A}, \textit{b}= 10.0284(19) \r{A} and \textit{c}= 12.9684(24) \r{A} (\textit{see} detailed structural information of $\AlthrNd$ in Supplementary Materials Table \rd{S1}).

\begin{figure}[htbp]
    \centering
    \includegraphics[width=1\linewidth]{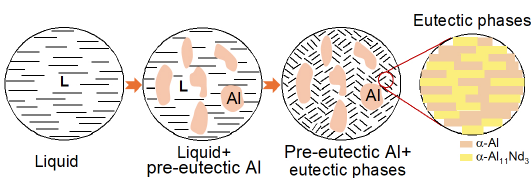}
    \caption{\textbf{Illustration of formation mechanism of the microstructure of the Al-3wt\%Nd alloy.}}
    \label{fig:eutec}
\end{figure}

With the structure information of $\alpha$-$\AleNdt$ and $\alpha$-Al, the morphology and crystallographic features of the microstructure of the as-casted Al-3wt\%Nd alloy are characterized by using electron backscatter diffraction (EBSD) technique. Figs.~\ref{fig:microstru}\rd{f$_1$} and \rd{f$_2$} show the overlapped measured and simulated electron  backscattered diffraction (\textit{i.e.}, $Kikuchi$) patterns for $\alpha$-Al and $\alpha$-$\AleNdt$, respectively. A good match between measured and simulated $Kikuchi$ patterns guarantees the accuracy of EBSD measurements. Figs.~\ref{fig:microstru}\rd{g$_1$} to \rd{g$_4$} display, respectively, band contrast (BC), phase distribution (PD), stacked BC and PD, and crystallographic orientation distribution plotted by using the index of inverse pole figure (IPF) along the horizontal direction of the sample (X$_0$). From phase distribution, it is confirmed that both the Al matrix and the Al phase possess the fcc structure, and the enriched-Nd $\AleNdt$ phase is of orthorhombic structure ($\alpha$-$\AleNdt$). Misorientation analyses show that no specific orientation relationship between the eutectic $\alpha$-$\AleNdt$ and $\alpha$-Al phases. Fig.~\ref{fig:eutec} illustrates the formation mechanism of the microstructure of the Al-3wt\%Nd alloy. With the cooling of temperature, when the solidification temperature is reached, the pre-eutectic $\alpha$-Al phase begins to crystallize and gradually grows from the liquid phase. When the eutectic point is reached, the remaining liquid simultaneously transforms into the eutectic $\alpha$-$\AleNdt$ and $\alpha$-Al phases.       

\subsubsection{Thermodynamic and elastic stability}
To clarify the stability of $\alpha$-$\AleNdt$ in the Al-3wt\%Nd alloy, \textit{ab-initio} calculations in the frame of density functional theory (DFT) are carried out. First, with the determined crystal structure information from the XRD refinement as input, a full structural relaxation is carried out. The optimized lattice constants of $\alpha$-$\AleNdt$ are $a$=4.3801(10) Å, $b$=10.0260(24) Å and $c$=13.0315(66) Å (\textit{see} detailed information in Supplementary Materials Section \rd{S2}). This result aligns well with the experimental results and previous reports~\cite{RN609, RN590} (Table~\ref{tab:latticeparameter}), suggesting the reliability of the adopted theoretical methods and computational parameters. With the optimized structural model, the thermomechanical stability of $\alpha$-$\AleNdt$ is evaluated. $E_{\mathrm{f}}$ of $\alpha$-$\AleNdt$ is determined to be $-$0.389 eV/atom. The negative value of $E_{\mathrm{f}}$ suggests a thermodynamic stability of $\alpha$-$\AleNdt$ against the decomposition into constituent elements. 

\begin{table}[htbp]
\centering
\caption{\textbf{Experimental and DFT-optimized lattice parameters of $\alpha$-$\AleNdt$.}}
\begin{tabular}{ccccc}
\cline{1-4}
Composition & \multicolumn{3}{c}{Lattice parameters (Å)} \\
            & \textit{a}     & \textit{b}     & \textit{c}      \\
\cline{1-4}
Present-Exp. & 4.3648  & 10.0284  & 12.9684         \\
Present-GGA & 4.3801(10)  & 10.0260(24)  & 13.0315(66)     \\
Exp.\textsuperscript{a}    & 4.36  & 10.02  & 12.93   \\
PBE-GGA\textsuperscript{b}    & 4.38  & 9.97 & 13.06  \\
\hline
\end{tabular}
\footnotesize
\raggedright \textsuperscript {a} Ref.~\cite{RN609}. \textsuperscript{b}  Ref.~\cite{RN590}
\label{tab:latticeparameter}
\end{table}

\begin{figure}[htbp]
    \centering
    \includegraphics[width=1\linewidth]{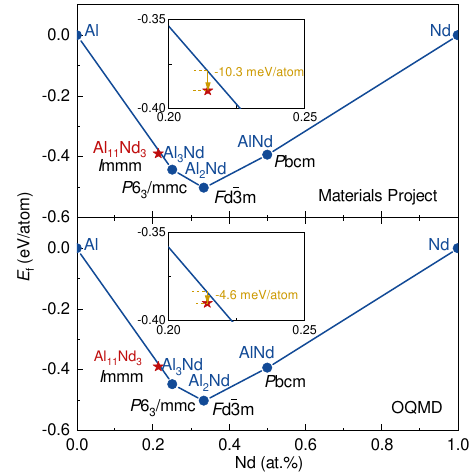}
    \caption{\textbf{Comparison of formation energy ($E_\mathrm{f}$) of $\alpha$-$\AleNdt$ with the other known convex-hull phases of binary Al-Nd system.} They include the fcc $\alpha$-Al (\textit{F}m$\overline{3}$m), the Ni$_3$Sn-type $\AlthrNd$ with D0$_{19}$ structure (\textit{P}6$_3$/mmc), the Cu$_2$Mg-type $\AltwNd$ with C15 laves structure (\textit{F}d$\overline{3}$m), the orthorhombic $\AlNd$ structure (\textit{P}bcm), and the $\alpha$-La-type Nd with hexagonal structure (\textit{P}6$_3$/mmc). Reference data in the upper and lower panels are obtained from the Materials Project \cite{Jain2013CommentaryTM} and OQMD~\cite{RN57, RN642, RN34} databases, respectively. The inset in each panel is the zoomed figure around $\AleNdt$.     
    }
    \label{fig:convhull}
\end{figure}

Fig.~\ref{fig:convhull} compares $E_{\mathrm{f}}$ of $\alpha$-$\AleNdt$ with the known convex-hull phases of the binary Al-Nd system~\cite{RN615, RN642, RN34, RN57}, including the Ni$_3$Sn-type $\AlthrNd$ with D019 structure (\textit{P}6$_3$/mmc), the Cu$_2$Mg-type $\AltwNd$ with C15 structure (\textit{F}d$\overline{3}$m), the orthorhombic $\AlNd$ structure (\textit{P}bcm). In the upper and lower panels, the referenced data is extracted from the Materials Project~\cite{Jain2013CommentaryTM} and OQMD~\cite{RN57, RN34, RN642} database, respectively. Clearly, $E_{\mathrm{f}}$ of $\alpha$-$\AleNdt$ lies at the outside of the connection line between $\alpha$-Al and $\AlthrNd$, as highlighted in the insets of Fig.~\ref{fig:convhull}. In the upper and lower panels, the energy distances of $\alpha$-$\AleNdt$ to the connection line between Al and $\AlthrNd$ are 10.3 and 4.6 meV/atom, respectively. Thus, $\alpha$-$\AleNdt$ would be a convex-hull phase and has a high thermomechanical stability against the decomposition into Al and $\AlthrNd$ (and other binary convex-hull phases). This result, aligning well with the previous report~\cite{RN590}, accounts for the experimental observation of $\alpha$-$\AleNdt$ in the Al-3wt\%Nd alloy.     

Based on the Born-Huang's lattice dynamics theory~\cite{RN579}, the elastic stability of $\alpha$-$\AleNdt$ is evaluated. First, using the strain-stress method~\cite{RN582}, the independent lattice constants ($\C{ij}$) of $\alpha$-$\AleNdt$ are determined.  Different from cubic crystals possessing three independent $\C{ij}$, the orthorhombic crystal has 9 independent $\C{ij}$, namely, $\C{11}$, $\C{12}$, $\C{13}$, $\C{22}$, $\C{23}$, $\C{33}$, $\C{44}$, $\C{55}$ and $\C{66}$~\cite{RN630}. Table~\ref{tab:elastic} lists the determined independent $\C{ij}$ of $\alpha$-$\AleNdt$ along with the one reported in the literature~\cite{RN595}. We see that the determined $\C{ij}$ is consistent with the previous report~\cite{RN595}. According to the Born-Huang's lattice dynamics theory~\cite{RN579}, the elastic stable orthorhombic system should satisfy the following three conditions simultaneously~\cite{RN630}:
\begin{equation}
  \C{ii} \textgreater 0 \, (i=1,\,4,\,5\ \mathrm{and}\,6),
\end{equation}
\begin{equation}
  \Delta_1=\C{11} \C{12} - \ \C{12}^2 \textgreater 0,
\end{equation}
\begin{align} 
  \Delta_2 =& \C{11}\C{22}\C{33} + 2\C{12}\C{13}\C{23} - \C{11}{\C{23}}^2  \nonumber \\
            & -\C{22}{\C{13}}^2-\C{33}{\C{12}}^2 \textgreater 0 
\end{align} 
For $\alpha$-$\AleNdt$, all $\C{ii}\, (i=1,\,4,\,5,\mathrm{and}\,6)$ are larger than 0, in which $C_{55}$ is the minimum one with a value of 42.3 GPa. Thus, the first condition is satisfied. For the second and third conditions, $\Delta_1$ and $\Delta_2$ are determined to be 0.37$\times$10$^4$ GPa$^2$ and 0.95$\times$10$^6$ GPa$^3$, respectively. Clearly, both values are larger than 0 significantly, suggesting the satisfaction of these two conditions simultaneously. The above results evidence a high elastic stability of $\alpha$-$\AleNdt$. 

\begin{table*}
\centering
\caption{\textbf{Independent elastic constants ($\C{ij}$) of $\alpha$-$\AleNdt$.} The orthorhombic crystal possesses nine independent elastic constants, namely, $\C{11}$, $\C {12}$, $\C{13}$, $\C{22}$, $\C{23}$, $\C{33}$, $\C{44}$, $\C{55}$ and $\C{66}$. For comparison, $\C{ij}$ of $\alpha$-$\AleNdt$, cubic Ni$_3$Al-type ($P$m$\overline{3}$m) and hexagonal Ni$_3$Sn-type ($P$6$_3$/mmc) $\AlthrNd$, cubic Cu$_2$Mg-type $\AltwNd$ ($F$d$\overline{3}$m) and cubic CsCl-type $\AlNd$ ($P$m$\overline{3}$m) reported in the literature~\cite{RN595,RN591,RN619} are also listed.}
\setlength{\tabcolsep}{2pt}
\begin{tabular*}{\linewidth}{cccccccccccc}
\cline{1-12}
Composition &           & Space Group  & $\C{11}$/GPa   & $\C{12}$/GPa  & $\C{13}$/GPa  & $\C{22}$/GPa  & $\C{23}$/GPa  & $\C{33}$/GPa  & $\C{44}$/GPa  & $\C{55}$/GPa  & $\C{66}$/GPa  \\
\cline{1-12}
\multirow{2}*{$\AleNdt$} & Present  & $I$mmm & 124.9 & 49.0 & 52.6 & 118.0 & 51.2 & 108.5 & 57.9  & 42.3  & 53.0    \\
            & PBE-GGA\textsuperscript{a} & $I$mmm  & 129.1   & 48.6    & 57.2    & 117.2   & 57.7    & 121.3   & 45.2    & 57.5    & 55.1    \\
$\AlthrNd$       & PBE-GGA \textsuperscript{b} & $P$m$\overline{3}$m      & 147.7  & 30.6  & $-$  & $-$  & $-$  & $-$  & 55.0   & $-$      & $-$   \\
$\AlthrNd$     & PBE-GGA \textsuperscript{c} & $P$6$_3$/mmc              & 76.3    & 32.7    & 30.9    & $-$      & $-$      & 256.6   & 70.2   & $-$   & 21.82    \\
$\AltwNd$       & PBE-GGA \textsuperscript{b} & $F$d$\overline{3}$m      & 149.3   & 34.51   & $-$   & $-$    & $-$      & $-$        & 45.46    & $-$      & $-$    \\
$\AlNd$        & PBE-GGA \textsuperscript{b} & $P$m$\overline{3}$m      & 68.9    & 51.3  & $-$    & $-$    & $-$   & $-$    & 49.8  & $-$      & $-$  \\
\hline    
\end{tabular*}
\footnotesize
\raggedright  \textsuperscript{a} Ref.~\cite{RN595}. \textsuperscript{b}  Ref.~\cite{RN591}.  \textsuperscript{c}  Ref.~\cite{RN619}
\label{tab:elastic}
\end{table*}

\subsection{Deformability and intrinsic material properties}\label{sec:dfmblt}

\subsubsection{Deformability during cold rolling}
\begin{figure}
\centering
    \includegraphics[width=1\linewidth]{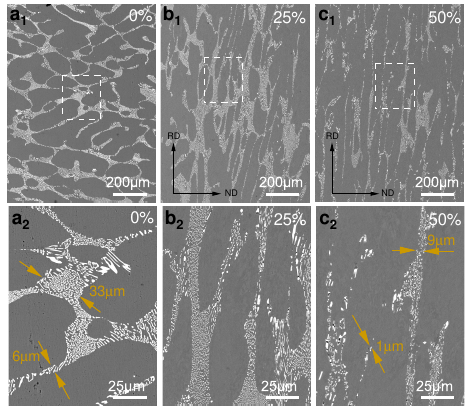}
    \caption{\textbf{Microstructure evaluation during asymmetric cold rolling.} (\rd{a$_1$}) 0\%. (\rd{b$_1$}) 25\%. (\rd{c$_1$}) 50\%. (\rd{a$_2$}-\rd{c$_2$}) are the zoomed figures highlighted in \rd{a$_1$}-\rd{c$_1$}, respectively. RD and ND represent the rolling and normal direction, respectively.}
    \label{fig:coldrolling}
\end{figure}

Combining experimental and \textit{ab-initio} theoretical study, the deformability of the Al-3wt\%Nd alloy and the intrinsic mechanical properties of $\alpha$-$\AleNdt$ are studied. First, the microstructure evolution of the Al-3wt\%Nd alloy during the asymmetrical cold rolling with a speed ratio of 1.3 is investigated. The selection of asymmetrical rolling aims to increase the deformability of metallic materials~\cite{vincze2020asymmetrical}. Figs.~\ref{fig:coldrolling}\rd{a$_1$}, \rd{b$_1$} and \rd{c$_1$} show the backscattered electron (BSE) images of the Al-3wt\%Nd alloy at the deformations of 0\%, 25\% and 50\%, respectively. Figs.~\ref{fig:coldrolling}\rd{a$_2$}, \rd{b$_2$} and \rd{c$_2$} are the zoomed images outlined by the dashed boxes in Figs.~\ref{fig:coldrolling}\rd{a$_1$}, \rd{b$_1$} and \rd{c$_1$}, respectively. Clearly, with the plastic deformation, the $\alpha$-Al matrix is obviously elongated along the rolling direction (RD, vertical direction in Fig.~\ref{fig:coldrolling}), aligning well with the excellent inherent ductility of Al and the sufficient dislocation slip systems of fcc metal. Accompanied by the elongation of $\alpha$-Al matrix, the eutectic microstructure is reshaped significantly. With the increased deformation, the cellular eutectic microstructure gradually turns to a lamellar shape with the lamella normal along the normal direction of the sample (ND, horizontal direction in Fig.~\ref{fig:coldrolling}).

At the deformation of 50\%, the thickness of the eutectic microstructure is measured to be 1$\sim$9 $\mu$m (Fig.~\ref{fig:coldrolling}\rd{c$_2$}), which is much smaller than that of the undeformed one (6$\sim$35 $\mu$m, Fig.~\ref{fig:coldrolling}\rd{a$_2$}). This result evidences the feasibility of homogenizing the microstructure of the Al-3wt\%Nd alloy by severe plastic deformation. Nevertheless,  despite a significant redistribution of the eutectic microstructure, the morphology and the dimensions of the eutectic $\alpha$-$\AleNdt$ phase do not change significantly.  After the deformation of 50\%, $\alpha$-$\AleNdt$ is still in a rod shape. The radius and lengths of $\alpha$-$\AleNdt$ are measured to be 0.4$\sim$4 $\mu$m and 0.6$\sim$15 $\mu$m, respectively. These values are very close to those of the as-casted ones (0.5$\sim$4 $\mu$m for the radius and 0.6$\sim$19 $\mu$m for the length). This result implies a difficulty of plastic deformation and fragmentation of  $\alpha$-$\AleNdt$ during cold rolling. 

\subsubsection{Intrinsic strength and brittleness-ductility}

\begin{figure}
    \centering
    \includegraphics[width=1\linewidth]{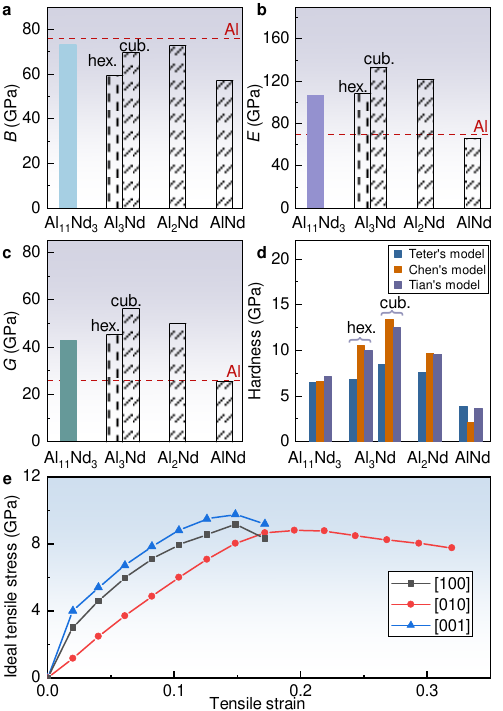}
    \caption{\textbf{Intrinsic mechanical strength of $\alpha$-$\AleNdt$.}  (\rd{a}) Bulk modulus ($B$). (\rd{b}) Young's modulus ($E$). (\rd{c}) Shear modulus ($G$). (\rd{d}) Hardness. (\rd{e}) Ideal stress-strain curves along $a$-, $b$- and $c$-axes. The dashed lines represent the pure $\alpha$-Al. Data of cubic (cub.) Ni$_3$Al-type~\cite{RN591} and hexagonal (hex.) Ni$_3$Sn-type~\cite{RN618} $\AlthrNd$, cubic Cu$_2$Mg-type $\AltwNd$~\cite{RN591} and cubic CsCl-type $\AlNd$~\cite{RN591} are also included as comparison. For hardness, the models proposed by D.M. Teter~\cite{RN634}, X.-Q. Chen~\cite{RN635} and Y. Tian~\cite{RN633} are adopted. }
    \label{fig:stren}
\end{figure}

\textit{a.\  Elastic moduli, hardness and ideal tensile strength} 
\ To understand the deformability of $\alpha$-$\AleNdt$, its intrinsic mechanical properties, including elastic moduli, hardess, ideal tensile strength and brittleness-ductility, are calculated. Figs.~\ref{fig:stren}\rd{a}, \rd{b} and \rd{c} show, respectively, the estimated polycrystalline bulk modulus ($B$), Young's modulus ($E$) and shear modulus ($G$)  of $\alpha$-$\AleNdt$ along with those of pure $\alpha$-Al as a reference. As a comparison, the elastic moduli of $\AlthrNd$, $\AltwNd$ and $\AlNd$ are also plotted~\cite{RN619,RN591,RN595}. Here, the Voigt-Reuss-Hill (VRH) averaging approach~\cite{RN630} (\textit{see} calculation details in Appendix) is adopted. We see that $B$ of $\alpha$-$\AleNdt$ (72.9 GPa) is comparable to that of pure $\alpha$-Al (76 GPa). On the contrary, $E$ (107 GPa) and $G$ (42.6 GPa) of $\alpha$-$\AleNdt$ are much higher than those of pure $\alpha$-Al, \textit{i.e.}, $E$=70 GPa and $G$=26 GPa~\cite{samsonov2012}. Furthermore, with the models proposed by D.M. Teter~\cite{RN634}, X.-Q. Chen~\cite{RN635} and Y. Tian~\cite{RN633}, the hardess of $\alpha$-$\AleNdt$ is estimated, as shown in Fig.~\ref{fig:stren}\rd{d}. For all models, the hardness of $\alpha$-$\AleNdt$ is estimated to be around 6.0 GPa (Fig.~\ref{fig:stren}\rd{d}), which is significantly larger than those of pure $\alpha$-Al, \textit{i.e.}, 160-350 MPa in Vickers hardness and 160-550 MPa in Brinell hardness. Compared with other Al-Nd binary compounds, it is found that the intrinsic mechanical strength of $\alpha$-$\AleNdt$ could be weaker than those of $\AlthrNd$ and $\AltwNd$, but is stronger than that of $\AlNd$, even though their differences are not prominent. 

The chemical bond-dominated ideal tensile strengths, which represent the maximum stress at elastic instability (yield or break) when applying an increased stress to an infinite defect-free crystal~\cite{morris2002,li1999}, are evaluated. Fig.~\ref{fig:stren}\rd{e} displays the ideal tensile stress-strain curves of $\alpha$-$\AleNdt$ along the $a$-, $b$- and $c$-axes of the orthorhombic cell.  During the calculation, the lattice is first deformed gradually along the tensile direction. Then, at each strain, the structure is relaxed until all stresses perpendicular to the tensile direction disappear with a threshold of Hellman-Feynman stress less than 0.1 GPa~\cite{RN564, RN562, RN563}.  The ideal tensile strains ($\varepsilon_\mathrm{max}$) along  the $a$-, $b$- and $c$-axes of $\alpha$-$\AleNdt$  are measured to be 14\%, 19\%, and 14\%, respectively. These values are much smaller than $\varepsilon_\mathrm{max}$ of pure $\alpha$-Al along [100] (36\%~\cite{li1999}). The ideal tensile stress ($\sigma_\mathrm{max}$) along the $a$-, $b$- and $c$-axes are determined to be 8.8 GPa, 8.3 GPa, and 9.2 GPa, respectively, which are also lower than that of pure $\alpha$-Al along [100] (12.5 GPa~\cite{li1999}). Both smaller $\varepsilon_\mathrm{max}$ and lower $\sigma_\mathrm{max}$ of $\alpha$-$\AleNdt$ against pure $\alpha$-Al imply a brittle nature of this compound. 

\begin{figure*}
    \centering
    \includegraphics[width=0.80\linewidth]{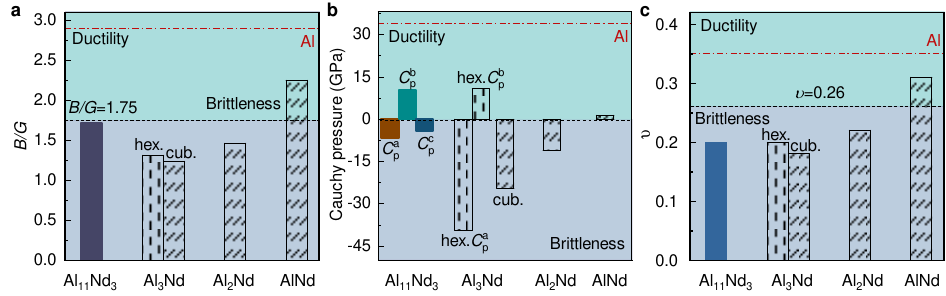}
    \caption{\textbf{Intrinsic ductility-brittleness of $\alpha$-$\AleNdt$.} (\rd{a}) Pugh’s ratio ($B/G$); (\rd{b}) Cauchy pressure ($C_\mathrm{p}$); (\rd{c}) Poisson’s ratio (\textit{v}). As a reference,  the data of cubic (cub.) Ni$_3$Al-type ~\cite{RN618} and hexagonal (hex.) Ni$_3$Sn-type ~\cite{RN618} $\AlthrNd$, cubic Cu$_2$Mg-type $\AltwNd$ ~\cite{RN618} and cubic CsCl-type $\AlNd$ ~\cite{RN591} are also plotted. The dash-doted line represents the pure $\alpha$-Al. The dashed line highlights the ductility-brittleness critical values, \textit{i.e.}, $B/G$=1.75 in Pugh's theory, $C_\mathrm{p}$=0 in Pettifor's model and $\nu$= 0.26 in Poisson's picture~\cite{RN563,RN650}. Note that for $C_\mathrm{p}$ of orthorhombic and hexagonal crystals, there are three ($C_\mathrm{p}^\mathrm{a}$, $C_\mathrm{p}^\mathrm{b}$ and $C_\mathrm{p}^\mathrm{c}$) and two ($C_\mathrm{p}^\mathrm{a}$ and $C_\mathrm{p}^\mathrm{b}$) components, respectively.}
    \label{fig:dctlbrtnes}
\end{figure*}
 
\textit{b.\ Inherent brittleness-ductility}
\ By using Pugh's, Pettifor's, and Poisson's criteria, the intrinsic brittleness-ductility of $\alpha$-$\AleNdt$ are quantitatively examined. In these three criteria, the brittleness-ductility are represented by $B/G$, Cauchy pressure $C_\mathrm{p}$, and Poisson's ratio $\nu$ (defined by $E/2G-1$~\cite{RN466}), respectively. As comparisons, the brittleness-ductilities of other pure $\alpha$-Al and other binary Al-Nd compounds are also evaluated. Note that unlike the cubic crystal having one component of $C_\mathrm{p}$ (defined by $\C{12}-\C{44}$), for the orthorhombic ($\alpha$-$\AleNdt$) and the hexagonal ($\AlthrNd$) crystals, their $C_\mathrm{p}$ possess three (\textit{i.e.}, $C_\mathrm{p}^\mathrm{a}$=$\C{23} -\C{44}$, $C_\mathrm{p}^\mathrm{b}$=$\C{13} -\C{55}$  and $C_\mathrm{p}^\mathrm{c}$=$\C{12} -\C{66}$~\cite{RN458,RN665,RN666}) and two (\textit{i.e.}, $C_\mathrm{p}^\mathrm{a}$=$\C{13} -\C{44}$ and $C_\mathrm{p}^\mathrm{b}$=$\C{12} -\C{66}$~\cite{RN665,RN666}) components, respectively.

Figs.~\ref{fig:dctlbrtnes}\rd{a}, \rd{b} and \rd{c} show, respectively, the determined $B/G$, $C_\mathrm{p}$, and $\nu$ of $\alpha$-$\AleNdt$ along with those of pure $\alpha$-Al (dash-dot lines), $\AlthrNd$, $\AltwNd$ and $\AlNd$. For clarity, the critical values of ductility-brittleness in different criteria, namely, $B/G$=1.75 in Pugh's theory, $C_\mathrm{p}$=0 in Pettifor's model and $\nu$= 0.26 in Poisson's picture~\cite{RN563,RN650}, are indicated by dashed lines. Clearly, the pure $\alpha$-Al with $B$/$G$ of 2.9, $C_\mathrm{p}$ of 33.9 GPa and $\nu$ of 0.35 is inherent ductility. For $\alpha$-$\AleNdt$, $B/G$ and $\nu$ are calculated to be 1.7 and 0.26, respectively. Clearly, both of them are located at the brittle regions in Pugh's (Fig.~\ref{fig:dctlbrtnes}\rd{a}) and Poisson's (Fig.~\ref{fig:dctlbrtnes}\rd{c}) criteria. The Cauchy pressure $C_\mathrm{p}^\mathrm{a}$, $C_\mathrm{p}^\mathrm{b}$ and $C_\mathrm{p}^\mathrm{c}$ are determined to be $-$6.7 GPa, 10.3 GPa and $-$4.1 GPa, respectively (Fig.~\ref{fig:dctlbrtnes}\rd{b}). The large deviation between various Cauchy pressure parameters implies a strong anisotropy of chemical bonding. Owing to the presence of two negative $C_\mathrm{p}$ parameters, $\alpha$-$\AleNdt$ should be considered inherently brittle from Pettifor's criteria. Despite the brittle nature, from all Pugh's, Pettifor's, and Poisson's criteria, the ductility of $\alpha$-$\AleNdt$ should be superior to those of $\AlthrNd$ and $\AltwNd$. On the contrary, it is inferior to the B2-type $\AlNd$, which should be classified as ductile since $B/G$, $C_\mathrm{p}$ and $\nu$ are larger than 1.75, 0 and 0.26, respectively. The ductility nature of B2-type $\AlNd$ aligns well with the result of B2-type AlSc~\cite{wang2023}. The above quantitative investigation evidences the hard and brittle nature of $\alpha$-$\AleNdt$, explaining well the difficulty in deformation and fragmentation of this phase in Al-3wt\%Nd alloy during plastic deformation. 

\subsection{Electronic structures}\label{sec:elecstru}
To understand the inherent mechanical properties of $\alpha$-$\AleNdt$, the electronic structures and chemical bonds of this compound are investigated systematically. First, the local chemical environments (LCEs), which decides electron structure and chemical bonding, are studied. Second, the electron band structure (BS) and density of states (DOS) are calculated. Third, the atom charge, quantum theory of atoms in molecular (QTAIM), crystal orbital Hamilton population (COHP)~\cite{RN577, RN34} and crystal orbital bond index (COBI)~\cite{muller2021crystal1} are investigated.

\subsubsection{Local chemical environment}
From the crystal structure of $\alpha$-$\AleNdt$, it is found that there are two inequivalent Nd atoms, occupying \textit{2a} (\Nd{I}) and \textit{4i} (\Nd{II}) sites, respectively, and four inequivalent Al atoms, located at \textit{2d} (\Al{I}), \textit{4h} (\Al{II}) and two \textit{8l} (\Al{III} and \Al{IV}) sites, respectively (\textit{see} details in Supplementary Materials Table \rd{S1}). Notably, in the unit cell of $\alpha$-$\AleNdt$, the numbers of two kinds of Nd atoms and four kinds of Al atoms are not the same, which can be seen from the multiplicity factor in the Wyckoff symbol. Specifically, in the unit cell, the number of \Nd{II} (\textit{4i}) is two times larger than that of \Nd{I} (\textit{2a}), and the numbers of \Al{III} and \Al{IV} (\textit{8l})  are two times and four times larger than \Al{II} (\textit{4h}) and \Al{I} (\textit{2d}), respectively. 

\begin{figure}
    \centering
    \includegraphics[width=1\linewidth]{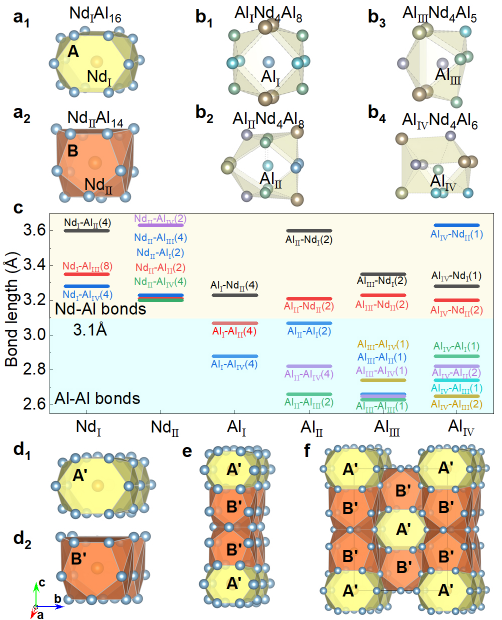}
    \caption{\textbf{Local chemical environment (LCE) in $\alpha$-$\AleNdt$.} (\rd{a$_1$}-\rd{a$_2$}) LCEs of \Nd{I} and \Nd{II}. (\rd{b$_1$}-\rd{b$_4$}) LCEs of \Al{I} to \Al{IV}. (\rd{c}) Distribution of bond length around distinct Nd and Al atoms. All Nd-Al bonds are longer than 3.1 \r{A}, whereas all Al-Al bonds are shorter than 3.1 \r{A}. The number in parentheses in the annotation represents the bond number. For instance, Nd\textsubscript{I}-Al\textsubscript{II}(4) indicate that there exists four Nd\textsubscript{I}-Al\textsubscript{II} bonds. (\rd{d$_1$}-\rd{d$_2$}) Column A' and B' polyhedron units constructed by stacking A and B polyhedrons along the \textit{a}-axis. (\rd{e}) Planar-shaped A'B'B'A' substructure constructed by connecting one A' and two B' columns alternately along the \textit{c}-axis; (\rd{f}) Structural model by stacking A'B'B'A' substructures along the \textit{b}-axis. Note that the two adjacent A'B'B'A' planar-shaped substructures are mutually shifted 1/2 lattice length along the \textit{a} and \textit{c} axes simultaneously.} 
    \label{localenv}
\end{figure}

\begin{figure*}\label{htbp}
    \centering
    \includegraphics[width=16cm]{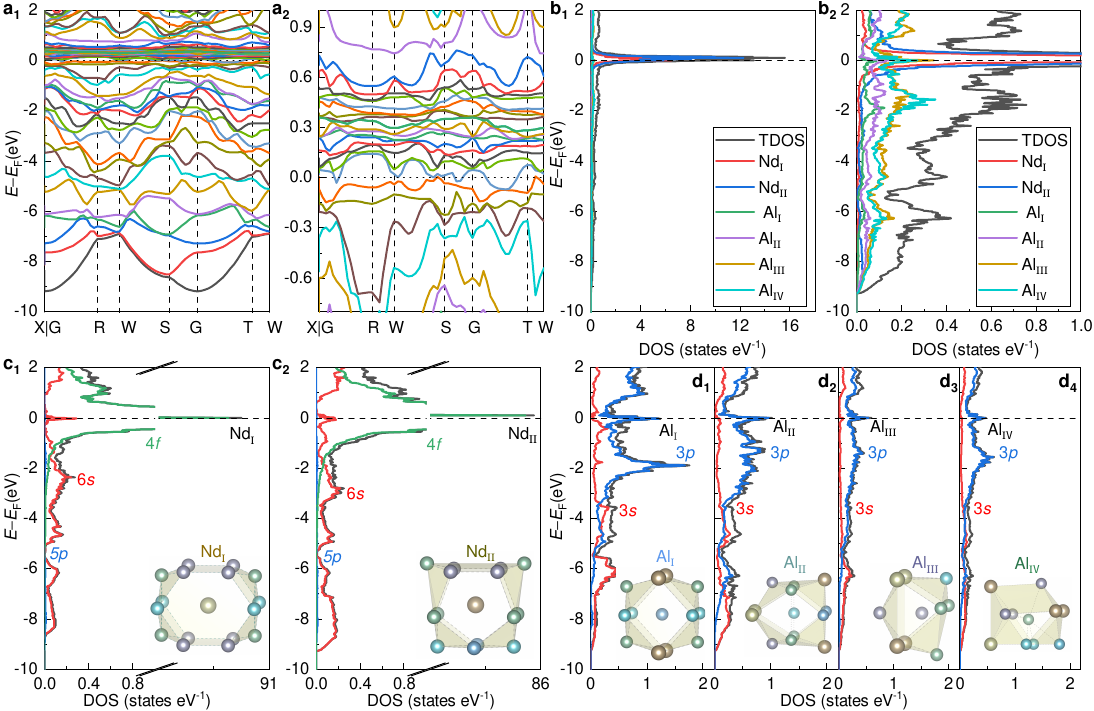}
    \caption{\textbf{Electronic band structure (BS) and density of states (DOSs) of $\alpha$-$\AleNdt$.}  (\rd{a$_1$}) BS along the high-symmetry points of the first Brillouin zone.  (\rd{a$_2$}) Zoomed BS around the Fermi energy ($E_\mathrm{F}$). (\rd{b$_1$}) Total DOS (TDOS) and atom-resolved partial DOS (PDOS). (\rd{b$_2$}) Zoomed DOS at the low-density region. (\rd{c$_1$}-\rd{c$_2$}) Atom and orbital-resolved DOSs of Nd$_\mathrm{I}$ and Nd$_\mathrm{II}$. (\rd{d$_1$}-\rd{d$_4$}) Atom and orbital-resolved DOSs of Al$_\mathrm{I}$, Al$_\mathrm{II}$, Al$_\mathrm{III}$, and Al$_\mathrm{IV}$.}    
    \label{fig:bsdos}
\end{figure*}

Figs.~\ref{localenv}\rd{a} and \rd{b} illustrate the local chemical environments of the two distinct Nd and four independent Al atoms, respectively. The distributions of bond lengths around different atoms are summarized in Fig.~\ref{localenv}\rd{c}. \Nd{I} is bonded with 16 Al atoms to form distorted \Nd{I}Al$_{16}$ cuboctahedra (Fig.~\ref{localenv}\rd{a$_1$}), including four \Al{II} with a length of 3.60 \r{A}, four \Al{IV} with a length of 3.28 \r{A}, and eight \Al{III} atoms with a length of 3.35 \r{A} (Fig.~\ref{localenv}\rd{c}).  \Nd{II} is bonded in a 12-coordinate geometry (Fig.~\ref{localenv}\rd{a$_2$}) to form distorted \Nd{II}Al$_{14}$ cuboctahedra, including two \Al{I} with a length of 3.23 \r{A}, two \Al{II} with a length of 3.21 \r{A}, four \Al{III} with a length of 3.23 \r{A}, four \Al{IV} with a length of 3.20 \r{A} and two \Al{IV} with a length of 3.63 \r{A} (Fig.~\ref{localenv}\rd{c}). Notably, all Nd atoms are bonded with Al atoms with bond lengths larger than 3.1 \r{A}, and no Nd-Nd bond exists. Like the Nd atoms, four kinds of Al atoms, \textit{i.e.}, \Al{I} (Fig.~\ref{localenv}\rd{b$_1$}), \Al{II} (Fig.~\ref{localenv}\rd{b$_2$}), \Al{III} (Fig.~\ref{localenv}\rd{b$_3$}) and \Al{IV} (Fig.~\ref{localenv}\rd{b$_4$}), possess different local chemical environments. Apart from the Al-Nd bond, some Al-Al bonds around the Al atom are observed.  Notably, as shown in Fig.~\ref{localenv}\rd{c}, all the lengths of the Al-Al bonds (\textless3.1 \r{A}) are shorter than those of the Al-Nd bonds (\textgreater3.1\r{A}).  

Remarkably, the crystal of $\alpha$-$\AleNdt$ can be constructed by stacking the polyhedrons of \Nd{I} (\Nd{I}Al$_{16}$, Fig.~\ref{localenv}\rd{a$_1$}) and \Nd{II} (Nd{II}Al$_{14}$, Fig.~\ref{localenv}\rd{a$_2$}) alternately. For easy illustration, the polyhedrons of \Nd{I}Al$_{16}$ and \Nd{II}Al$_{14}$ are dubbed A and B, respectively.  First, by stacking the polyhedron A and B along the \textit{a}-axis, the column A' (Fig.~\ref{localenv}\rd{d$_1$}) and B' (Fig.~\ref{localenv}\rd{d$_2$}) polyhedron units can be obtained, respectively.  Second, by connecting two A' and two B' column units alternately along the \textit{c}-axis, the planar-shaped A'B'B'A' substructure along the \textit{ac} plane is formed (Fig.~\ref{localenv}\rd{e}). Third, the whole crystal can be constructed (Fig.~\ref{localenv}\rd{f}) by stacking A'B'B'A' substructures one by one along the \textit{b}-axis.  Notably, the two adjacent A'B'B'A' planar-shaped substructures are mutually shifted 1/2 lattice length along the \textit{a} and \textit{c} axes simultaneously.

\subsubsection{Band structure and density of states}\label{sec:dos}

Fig.~\ref{fig:bsdos}\rd{a$_1$} displays the electron band structure (BS) of $\alpha$-$\AleNdt$. In terms of dispersion degree, the BS of $\alpha$-$\AleNdt$ could be classified into three distinct regions. For the energies being lower and higher Fermi energy ($E_\mathrm{F}$), the bands show obvious dispersion along various high-symmetry \textit{k}-point paths. In contrast, the nearly dispersionless flat bands are observed around $E_\mathrm{F}$. From the zoomed BS around $E_\mathrm{F}$, it is seen that the flat bands span from $-$0.2 eV to 0.5 eV (Fig.~\ref{fig:bsdos}\rd{a$_2$}). Among them, two bands are located below $E_\mathrm{F}$, one is crossed with $E_\mathrm{F}$, and the others are above $E_\mathrm{F}$. Fig.~\ref{fig:bsdos}\rd{b$_1$} shows the total density of states (TDOS) and the atom-resolved partial density of states (PDOS) of $\alpha$-$\AleNdt$ with the low-density region highlighted in Fig.~\ref{fig:bsdos}\rd{b$_2$}. Remarkably, the states around $E_\mathrm{F}$ are much higher than those of lower and higher $E_\mathrm{F}$, aligning well with the dispersion feature of BS. Comparing TDOS and PDOS of different atoms, it is evident that the high-density states around $E_\mathrm{F}$ mostly come from \Nd{I} and \Nd{II}, while the low-density states being away from $E_\mathrm{F}$ are majorly associated with the \Al{I} to \Al{IV} atoms.
 
Figs.~\ref{fig:bsdos}\rd{c} and \rd{d} display the orbital-resolved DOS for the two independent Nd and the four distinct Al atoms, respectively. Unexpectedly, the atom and orbital-resolved DOS structures of \Nd{I} (Fig.~\ref{fig:bsdos}\rd{c$_1$}) and \Nd{II} (Fig.~\ref{fig:bsdos}\rd{c$_2$}) are very similar, even though they possess totally different local chemical environments. Clearly, the high-density states of Nd around $E_\mathrm{F}$ originate from 4\textit{f} electrons, in good line with the observation in other materials containing rare earth element~\cite{patthey1990high}. Besides, some low-density states (\textless0.2 states/eV) stemming from 6\textit{s} electrons of Nd are also observed. For the Al atoms (Figs.~\ref{fig:bsdos}\rd{d$_1$}-\rd{d$_4$}), there exist four obvious state peaks centered at $-$6 eV, $-$4 eV, $-$2 eV and $E_\mathrm{F}$, respectively. The peak centered at $-$6 eV majorly comes from 3\textit{s} electrons, the one centered at $-$4 eV is contributed by both 3\textit{s} and 3\textit{p} electrons, and the peak centered at $-$2 eV and $E_\mathrm{F}$ are mainly originated from 3\textit{p} electrons. Unlike the similar DOSs of \Nd{I} and \Nd{II}, there exist obvious discrepancies for the four kinds of Al atoms, suggesting a high sensitivity of the electron structure of Al with respect to the local chemical environment. From Figs.~\ref{fig:bsdos}\rd{c} and \rd{d}, there could exist the interaction between 6\textit{s} electron of Nd and 3\textit{s} electron of Al at energies around $-$6 eV and $-$4 eV and the interaction of 6\textit{s}/4\textit{f} electron of Nd and 3\textit{p} electron of Al around $E_\mathrm{F}$. However, owing to the weak density, these interactions between Al and Nd would be not strong. Comparing the PDOSs of four kinds of Al atoms, it is deduced that there could exist 3\textit{s}-3\textit{p} hybridization at energies centered $-$4 eV and 3\textit{p}-3\textit{p} interaction at energies centered $-$2 eV between two neighboring Al atoms. 

\begin{figure*}
    \centering
    \includegraphics[width=1\linewidth]{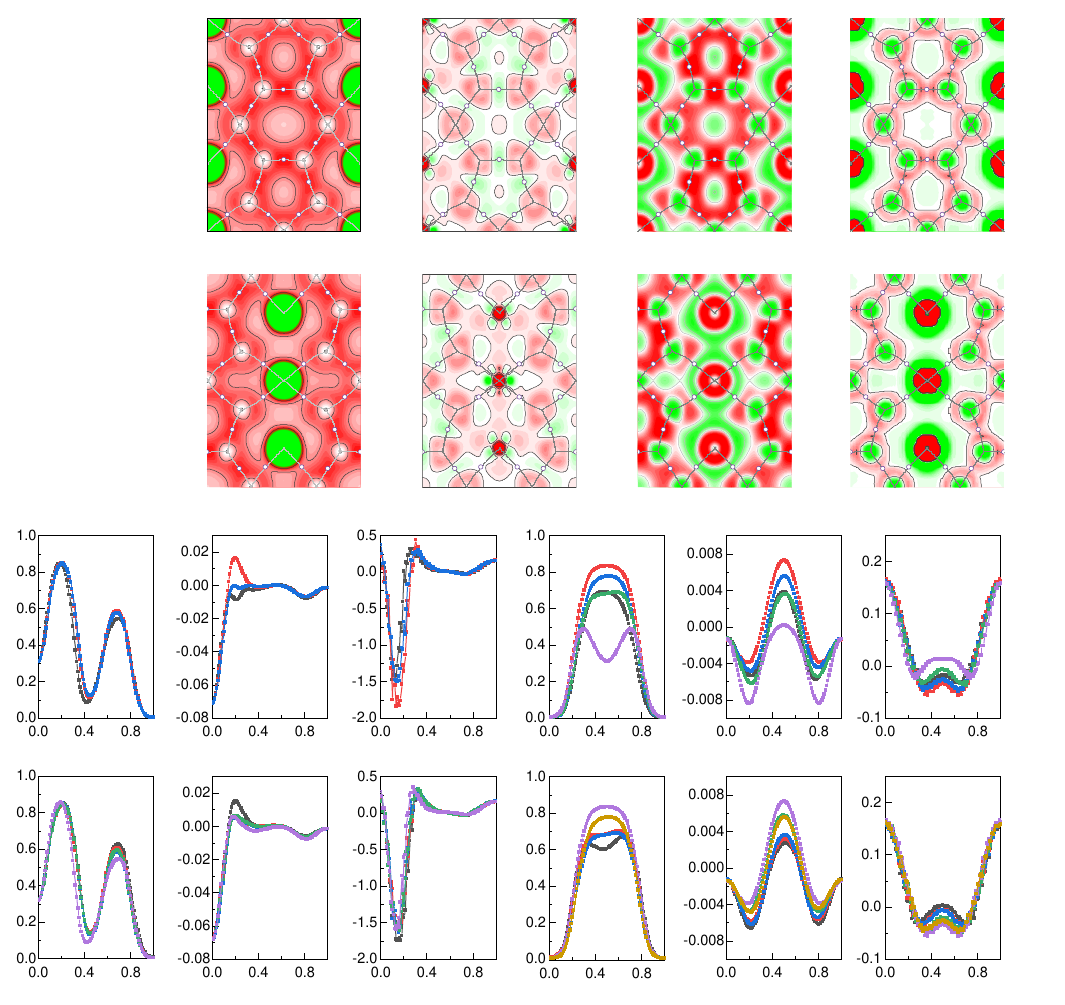}
    \caption{\textbf{Topological analysis of charge density.} (\rd{a$_1$}-\rd{b$_1$}) Atom arrangement on (1\,0\,0) and (0.5\,0\,0). (\rd{a$_2$}-\rd{b$_2$}) Charge density (CHG). (\rd{a$_3$}-\rd{b$_3$}) Charge density difference (CDD). (\rd{a$_4$}-\rd{b$_4$}) Electron localization function (ELF). (\rd{a$_5$}-\rd{b$_5$}) Laplacian of charge density (LAP). The grey lines and the open cycles represent the bond paths (BPs) and the bond critical points (BCPs), respectively. (\rd{c$_1$}-\rd{c$_3$}), (\rd{d$_1$}-\rd{d$_3$}), (\rd{e$_1$}-\rd{e$_3$}) and (\rd{f$_1$}-\rd{f$_3$}) are line-profiles of ELF, CDD and LAP along the bonds between \Nd{I} and its surrounding Al atoms, along the bonds between various Al atoms around \Nd{I}, along the bonds between \Nd{II} and its surrounding Al atoms and along the bonds between various Al atoms around \Nd{II}, respectively. The legend in \rd{c$_1$}-\rd{c$_3$}, \rd{d$_1$}-\rd{d$_3$}, \rd{e$_1$}-\rd{e$_3$} and \rd{f$_1$}-\rd{f$_3$} are the same, respectively. For a better comparison, the lengths of the bonds are normalized.}
    \label{fig:elfcdd}
\end{figure*}

\subsubsection{Chemical bonding}
\textit{a.\ Population analysis}
\ To reveal the features of chemical bonds in $\alpha$-$\AleNdt$, the atom charges are calculated. Table~\ref{tab:bader} lists the transferred charges of different Nd and Al atoms during the formation of chemical bonding calculated following the charge division algorithms proposed by Bader, Mulliken and Löwdin~\cite{RN604,RN570,RN569}. Despite the distinct charge values, the conclusion about charge transfer is consistent. Specifically, during the bonding, both \Nd{I} and \Nd{II} atoms lose electrons and all Al atoms get electrons, which is in accordance with the fact of more positive electronegativities of Nd (1.14 in the Pauling scale) compared with Al (1.61 in the Pauling scale). This result implies that the ionic bond could form between Nd and Al atoms in $\alpha$-$\AleNdt$. As shown in Table~\ref{tab:bader}, despite different local chemical environments, the transferred charges for \Nd{I} and \Nd{II} are similar, and so are they for all kinds of Al atoms (Table~\ref{tab:bader}). Thus, the strength of Coulomb electrostatic interaction between different Nd-Al pairs could be compared.   
 
\begin{table}[htbp]
\renewcommand{\arraystretch}{1.5}
\centering
\caption{\textbf{Charge transfer of Al and Nd during the formation of chemical bonding in $\alpha$-$\AleNdt$.}}
\setlength{\tabcolsep}{2pt}
\begin{tabular*}{\linewidth}{cccc}
\hline
Atom & Bader charge (\textit{e}) & Mulliken charge (\textit{e}) & Löwdin charge (\textit{e}) \\
\hline
        \Nd{I}   & $-$0.98 &$-$2.56 & $-$2.45    \\
         \Nd{II}  & $-$0.91  &$-$2.45 & $-$2.43   \\
        \Al{I}    & 0.14 & 0.57  & 0.69        \\
        \Al{II}   & 0.21 & 0.67  & 0.60       \\
         \Al{III}  & 0.32 & 0.73  & 0.72   \\
         \Al{IV}   &0.25 &0.66  & 0.64 \\         
\hline
\end{tabular*}
\label{tab:bader}
\end{table}

\textit{b.\ QTAIM and ELF topological analysis}
\ By using the quantum theory of atoms in molecular (QTAIM)~\cite{cortes2005}, the topological analyses on the charge density of $\alpha$-$\AleNdt$ are performed. In this work, the charge density (CHG, $\rho$(\textbf{r})), charge density difference (CDD, $\Delta\rho$(\textbf{r})=$\rho$(\textbf{r})$-$$\rho$(\textbf{r})$_\mathrm{unbonding}$) reflecting electron transfer during the bonding, electron localization function (ELF, $\eta$(\textbf{r})) characterizing electron localization with a range of [0, 1]~\cite{savin1997elf}  and Laplacian of charge density (LAP, $\nabla^2\rho$(\textbf{r})) determining local concentration or depletion of electron~\cite{macdougall1989}, are calculated. Figs.~\ref{fig:elfcdd}\rd{a} and \rd{b} show the 2-dimensional (2D) distributions of $\rho$(\textbf{r}), $\Delta\rho$(\textbf{r}), $\eta$(\textbf{r}) and  $\nabla^2\rho$(\textbf{r}) on (1\,0\,0) and (0.5\,0\,0) of $\alpha$-$\AleNdt$, respectively. The grey line represents the bond path, a single line of maximum electron density linking the nuclei of two chemically bonded atoms~\cite{bader1998bond, matta2007}. The open cycle indicates the bond critical points (BCP), a single point with minimum electron density along the bond path~\cite{cortes2005}. The values of $\rho$, $\Delta\rho$, $\eta$ and $\nabla^2\rho$ at the BCPs, which can reflect the bonding features~\cite{bader1998bond, matta2007}, are listed in Table~\ref{tab:lap}. Notably, there is no obvious charge accumulation between Nd and Al (Figs.~\ref{fig:elfcdd}\rd{a$_2$}, \rd{b$_2$}, \rd{a$_3$} and \rd{b$_3$}). At the BCPs of the Nd-Al bonds, the localization of electrons is very weak (Figs.~\ref{fig:elfcdd}\rd{a$_4$} and \rd{b$_4$}). The Laplacian of charge density exhibits positive values, suggesting a local depletion of electron, \textit{i.e.}, the electron density is less than the average density in the immediate neighborhood~\cite{aray2000laplacian}. At the BCPs of the Nd-Al bonds,  $\rho$, $\Delta\rho$, $\eta$ and  $\nabla^2\rho$ are around 0.02 $e$/bohr$^3$, 0.001 $e$/bohr$^3$, 0.2, and 0.02 $e$/bohr$^5$, respectively (Table~\ref{tab:lap}). All these features confirm the ionic-type nature of the Nd-Al bonds in $\alpha$-$\AleNdt$.

\begin{table}
    \caption{\textbf{Charge density ($\rho$), charge density difference ($\Delta\rho$), electron localization function ($\eta$), and Laplacian charge density ($\nabla^2\rho$) at bond critical points (BCPs) of different Al-Nd and Al-Al bonds.}}
    \centering
    \setlength{\tabcolsep}{5pt}
    \begin{tabular*}{\linewidth}{cccccc}
    \hline
        \multirow{2}{*}{Type} & Distance& $\rho$ & $\Delta\rho$ &\multirow{2}{*}{$\eta$} & $\nabla^2\rho$,\\
            &  Å & \textit{e}/bohr$^3$ &  \textit{e}/bohr$^3$ & & \textit{e}/bohr$^5$\\
        \hline
        \Nd{II}-\Al{II}   & 3.21  & 0.024	&0.0008   &  0.240  &  0.023   \\
        \Nd{II}-\Al{IV}   & 3.20  & 0.021	&0.0005  &0.224   & 0.025           \\
        \Nd{II}-\Al{I}    & 3.23  & 0.021  &0.00023  & 0.220   & 0.024       \\
        \Nd{II}-\Al{III}  & 3.23  & 0.021    & -0.00005  &0.210    &0.025        \\
        \Nd{I}-\Al{IV}    &3.28   & 0.020   &0.0002  &0.204  & 0.024   \\  
         \hline
        \Al{III}-\Al{III} &2.63   &  0.040	& 0.007  & 0.835    & $-$0.029   \\
        \Al{IV}-\Al{III}  & 2.65  &  0.038	& 0.006     &0.780   & $-$0.022  \\
        \Al{III}-\Al{II}  & 2.66  &  0.036	& 0.004   & 0.685   & $-$0.013    \\
        \Al{IV}-\Al{III}  & 2.74  & 0.036	&  0.007   & 0.833   &  $-$0.020   \\
        \Al{I} -\Al{IV}   & 2.88  & 0.030  & 0.004  &  0.727  & $-$0.006       \\
        \Al{II} -\Al{I}   &3.07   &0.025   &0.003  &  0.614  &  0.003    \\
        \hline
    \end{tabular*}
    \label{tab:lap}
\end{table}
   
Unlike the Nd-Al bond, as shown in Figs.~\ref{fig:elfcdd}\rd{a$_2$}, \rd{b$_2$}, \rd{a$_3$} and \rd{b$_3$}, obvious charge accumulation is observed between the Al and Al atoms. The $\rho$ at the BCPs of the Al-Al bonds is around 0.03-0.04 $e$/bohr$^3$, which is two times higher than that of the Nd-Al bond. From the distribution of ELF, the electrons between two neighboring Al atoms are found to possess obvious localization (Fig.~\ref{fig:elfcdd}\rd{a$_4$} and \rd{b$_4$}). The $\eta$ values at the BCPs of the Al-Al bonds are around 0.8. Furthermore, unlike the positive value of the Nd-Al bond, $\nabla^2\rho$ at the BCPs of the Al-Al bonds are negative, implying a local concentration of electrons. All these features suggest a covalent-type nature of the Al-Al bond in $\alpha$-$\AleNdt$.
 
To highlight the differences between various Nd-Al bonds and between various Al-Al bonds, the 1-dimensional (1D) line profiles of $\eta$(\textbf{r}), $\Delta\rho$(\textbf{r}) and $\nabla^2\rho$(\textbf{r}) are plotted. Figs.~\ref{fig:elfcdd}\rd{c}, \rd{d}, \rd{e} and \rd{f} show the line profiles of $\eta$(\textbf{r}), $\Delta\rho$(\textbf{r}) and $\nabla^2\rho$(\textbf{r}) along the bonds between \Nd{I} and its surrounding Al atoms, the bonds between various Al atoms around the \Nd{I} atom, the bonds between the \Nd{II} and its surrounding Al atoms, the bonds between various Al atoms around the \Nd{II} atom, respectively. For an easy comparison, the lengths of all bonds are normalized. From all $\eta$(\textbf{r}) (Figs.~\ref{fig:elfcdd}\rd{c$_1$} and \rd{e$_1$}), $\Delta\rho$(\textbf{r}) (Figs.~\ref{fig:elfcdd}\rd{c$_2$} and \rd{e$_2$}) and $\nabla^2\rho$(\textbf{r}) (Figs.~\ref{fig:elfcdd}\rd{c$_3$} and \rd{e$_3$}), all ionic-type Nd-Al bonds exhibit similar features, which aligns well with the similar atom charges of the Nd and Al atoms (Table~\ref{tab:bader}). This result shows that the ionic-type Nd-Al bond in $\alpha$-$\AleNdt$ is insensitive to bond length and local chemical environment. Unlike the Nd-Al bond, $\eta$(\textbf{r}), $\Delta\rho$(\textbf{r}) and $\nabla^2\rho$(\textbf{r}) of different Al-Al bonds exhibit obvious discrepancies, which is in agreement with the distinct DOS structures of the Al atoms (Fig.~\ref{fig:bsdos}\rd{d}). It shows that compared with the Al-Nd bond, the Al-Al bond is relatively sensitive to bond length and local environment. 

\begin{figure*}
    \centering
    \includegraphics[width=0.75\linewidth]{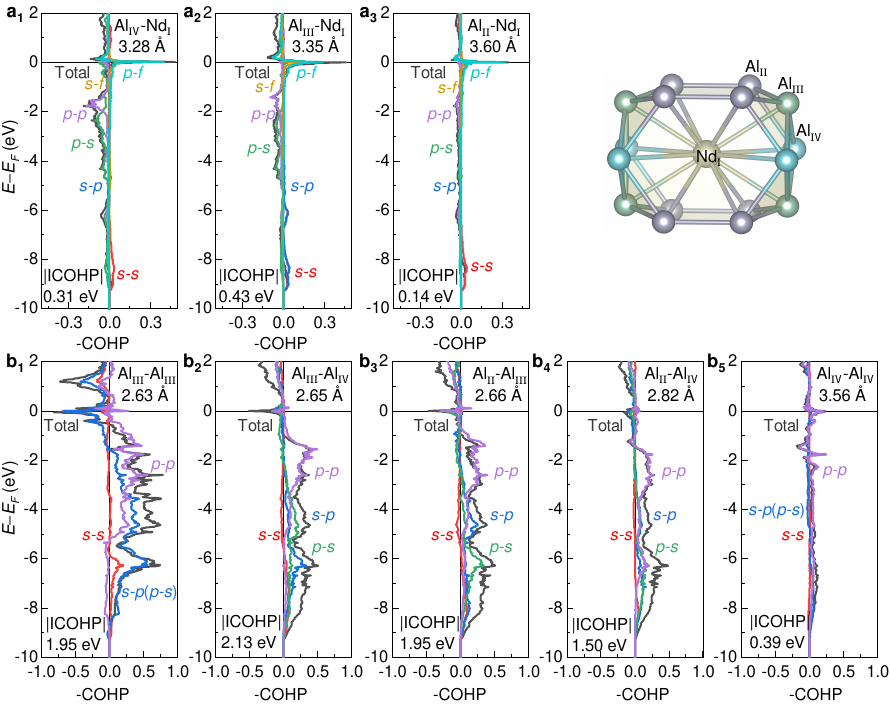}
    \caption{\textbf{COHP and orbital-resolved COHP curves of chemical bonds around \Nd{I}}. (\rd{a$_1$}$-$\rd{a$_3$}) The bonds between \Nd{I} and its surrounding Al atoms. (\rd{b$_1$}$-$\rd{b$_5$}) The bonds between different Al atoms around \Nd{I}. In both \rd{a} and \rd{b}, from left to right, the bond length increases gradually. }
    \label{fig:cohp1}
\end{figure*}

\begin{figure*}
    \centering
    \includegraphics[width=0.75\linewidth]{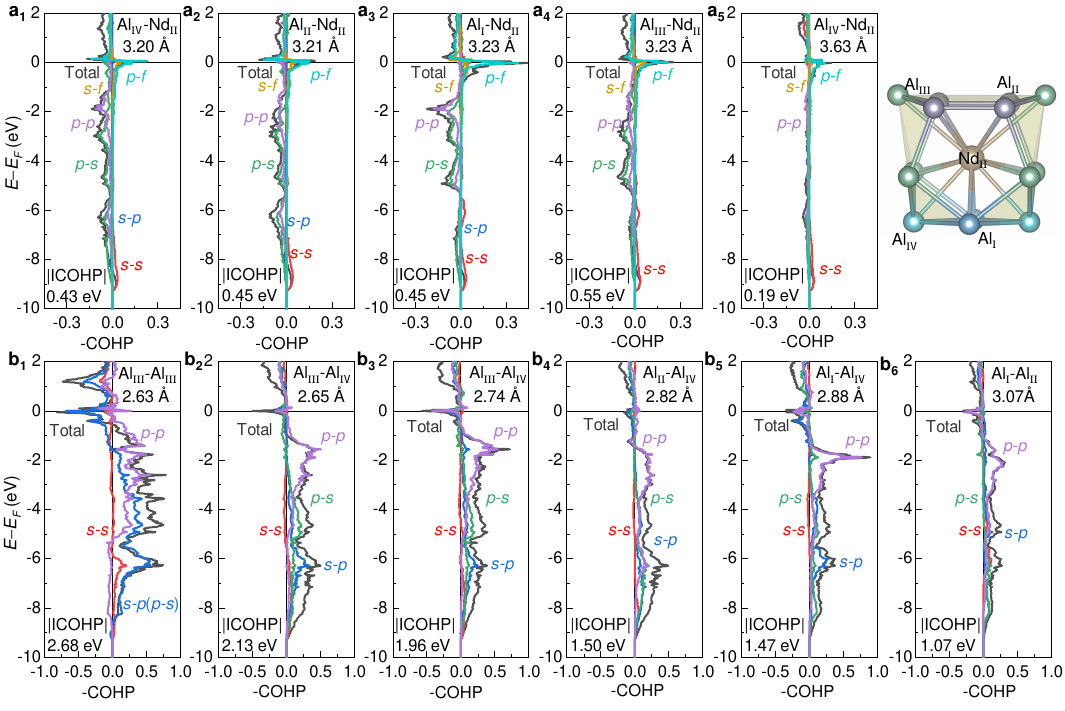}
    \caption{\textbf{COHP and orbital-resolved COHP curves of chemical bonds around \Nd{II}}. (\rd{a$_1$}$-$\rd{a$_5$}) The bonds between \Nd{II} and its surrounding Al atoms. (\rd{b$_1$}$-$\rd{b$_6$}) The bonds between different Al atoms around \Nd{II}. In both \rd{a} and \rd{b}, from left to right, the bond length increases gradually.}
    \label{fig:cohp2}
\end{figure*}

\textit{c.\ COHP, ICOHP and ICOBI}
\ Figs.~\ref{fig:cohp1} and~\ref{fig:cohp2}  display the crystal orbital Hamilton population (COHP) and the orbital-resolved COHP curves for the chemical bonds around \Nd{I} and \Nd{II}, respectively. The integration of total and orbital-resolved COHPs up to $E_\mathrm{F}$ (ICOHPs), an indicator of bonding strength~\cite{RN577,RN578}, are listed in Table~\ref{tab:Al-Nd_cohp}. Remarkably, the absolute values of ICOHP ($\mid$ICOHP$\mid$) of the Nd-Al bonds are generally smaller than 0.5 eV, while $\mid$ICOHP$\mid$ of the Al-Al bonds are around 1.5-2.5 eV. Higher bonding strength of the Al-Al bond may link with their shorter bond length (\textless3.1 \r{A}) with respect to the Nd-Al bond (\textgreater3.1 \r{A}).  For the Al-Al bond, as shown in Table~\ref{tab:Al-Nd_cohp}, the bonding strength decreases significantly with the increase in bond length. For instance, $\mid$ICOHP$\mid$ of \Al{III}-\Al{III} with a length of 2.63 \r{A} is 2.68 eV, while $\mid$ICOHP$\mid$ drops to 1.51 eV when the length increases to 2.82 \r{A} (\Al{II}-\Al{IV}). This result further evidences the high sensitivity of the Al-Al bonds on local environment.  
  
\begin{table}[htbp]
\renewcommand{\arraystretch}{1.5}
\centering
\caption{\textbf{Total and orbital-resolved integration of COHP (ICOHP) and the integration of crystal orbital bond index (ICOBI)~\cite{muller2021crystal1} for chemical bonds around \Nd{I} and \Nd{II}.} }
\label{tab:Al-Nd_cohp}
    \setlength{\tabcolsep}{2pt}
\begin{tabular*}{\linewidth}{cccccccccc}
\cline{1-10}
\multirow{2}{*}{Type}&\textit{d}&\multicolumn{7}{c}{$\mid$ICOHP$\mid$ (eV)} & \multirow{2}{*}{ICOBI}\\
  &  (Å)   & $s$-$s$    &$p$-$s$    & $s$-$p$    & $p$-$p$    & $s$-$f$     & $p$-$f$    & Total  \\
 \cline{1-10}
 {\Al{IV}-\Nd{I}}     &3.28      &0.058  &0.141    &0.005      &0.048  &0.018  &0.047  &0.307 &0.07\\
 {\Al{III}-\Nd{I}}    &3.35       &0.084	   &0.243	&0.005	&0.043	 &0.019	 &0.048	 &0.432 &0.10\\
 {\Al{II}-\Nd{I}}     &3.60       &0.027 	   &0.080 	&0.001 	&0  &0.006 	&0.030 	 &0.142 &0.05\\
 \cline{2-10}
 {\Al{III}-\Al{III}}  &2.63      &0.108	   &—   &1.417	&1.157	     &—      &—        &2.682 & 0.54\\
 {\Al{III}-\Al{IV}}   &2.65      &0.117      &— 	&1.170	&0.842       &—  &—        &2.130 &0.49\\
 {\Al{II}-\Al{III}}   &2.66      &0.080      &—	&1.040	&0.833	     &—   &—          &1.953 &0.45\\ 
 {\Al{II}-\Al{IV}}    &2.82      &0.010      &—	&0.800	&0.700      &—	   &—          &1.510 &0.31\\ 
 {\Al{IV}-\Al{IV}}    &3.56     &0.008       &—	&0.058 	&0.318 	&—	    &—         &0.384 &0.09\\ 
\cline{1-10}
{\Al{IV}-\Nd{II}}    &3.20    &0.061	 &0.219  &0.008	  &0.070	&0.026	&0.057	&0.425 &0.09\\
{\Al{II}-\Nd{II}}    &3.21    &0.052	 &0.259	&0.008	  &0.057	&0.028	&0.058	&0.446 &0.09\\
{\Al{I}-\Nd{II}}     &3.23    &0.094	 &0.2	&0.007	  &0.056	&0.034	&0.069	&0.446 &0.12\\
{\Al{III}-\Nd{II}}   &3.23    &0.093	 &0.307	&0.008	  &0.075	&0.024	&0.057	&0.548 &0.11\\
{\Al{IV}-\Nd{II}}    &3.63    &0.034  &0.110 	&0	      &0.005 	&0.007 	&0.038 	&0.194 &0.06 \\
\cline{2-10}
{\Al{III}-\Al{III}}  &2.63    &0.108	  &—       &1.417	 &1.157	&—      &—   &2.682 &0.54\\  
{\Al{III}-\Al{IV}}   &2.65    &0.117   &—	    &1.170   &0.842	&—    &—   &2.130 &0.49\\ 
{\Al{III}-\Al{IV}}   &2.74    &0.102	 &—       &1.081    &0.780	&—       &—   &1.963 &0.46\\ 
{\Al{II}-\Al{IV}}    &2.82    &0.010  &—		  &0.800	  &0.700   &—	    &—		&1.510 &0.46\\ 
{\Al{I}-\Al{IV}}     &2.88    &0.012  &—	      &0.841	   &0.614  &—        &—	     &1.467 &0.33\\
{\Al{I}-\Al{II}}     &3.07    &0.004  &—        &0.521    &0.541	&—          &—      &1.066 &0.25\\
\cline{1-10}
\end{tabular*}
\end{table} 

Comparing the total and orbital-resolved $\mid$ICOHP$\mid$, it is evident that the bonding strength of the Al-Al bond is governed by both 3\textit{s}-3\textit{p} and 3\textit{p}-3\textit{p} hybridization, aligning well with the DOS results (Section~\ref{sec:dos}). Therein, the former, \textit{i.e.}, 3\textit{s}-3\textit{p} interaction, plays a slightly larger contribution. From Figs.~\ref{fig:cohp1}\rd{b} and~\ref{fig:cohp2}\rd{b}, it is seen that the bonding states of 3\textit{s}-3\textit{p} interaction are located at the low-energy regions with the state peaks centered around $-$6 and $-$4 eV. For the Al-Al bonds with shorter distances, such as \Al{III}-\Al{III} bonds (Fig.~\ref{fig:cohp1}\rd{b$_1$} and Fig.~\ref{fig:cohp2}\rd{b$_1$}), some anti-bonding states of 3\textit{s}-3\textit{p} interaction are observed around $E_\mathrm{F}$. Nevertheless, these anti-bonding states disappear gradually with the increase in bond length. Compared with the 3\textit{s}-3\textit{p} interaction, the 3\textit{p}-3\textit{p} hybridization occurs at the energies being closer to $E_\mathrm{F}$. For all Al-Al bonds, an obvious bonding state peak centered around $-$2 eV is observed. Notably, with the increase in bond length, the intensity of the bonding state of both 3\textit{s}-3\textit{p} and 3\textit{p}-3\textit{p} interaction decreases significantly, which accounts for the high sensitivity of the bonding strength of Al-Al bonds. 

Lastly, the crystal orbital bond index (COBI) for different bonds of $\alpha$-$\AleNdt$ are calculated. Table~\ref{tab:Al-Nd_cohp} lists the integration of COBI up to $E_\mathrm{F}$ (ICOBI), which is an index of bond order in solid~\cite{muller2021crystal1}.  It is seen that the ICOBIs of all Nd-Al bonds are around 0.1, which is comparable to the Na-Cl bond (0.09~\cite{muller2021crystal1}) in  NaCl. This result evidences the strong ionic bonding nature of the Nd-Al bond. For the Al-Al bonds, ICOBIs are determined to be around 0.5, which is around half of C-C bond in diamond (0.95~\cite{muller2021crystal1}). Thus, despite the presence of electron localization between Al-Al, the covalencies of the Al-Al bonds in $\alpha$-$\AleNdt$ are much weaker than those of typical covalent compounds, such as diamond or silicon. This result is consistent with the metallic bonding of Al-Al bonding in pure Al. As is known, it is not easy to distinguish weak covalent interaction and metallic bonding in intermetallics~\cite{anderson1994metallic}. Therefore, it can be concluded that the Nd-Al is of typical ionic-type chemical bonding, while the Al-Al bonds around Nd possess weak-covalent or metallic bonds. The mixed ionic and weak-covalent bond is responsible for the hardness-brittleness of $\alpha$-$\AleNdt$.
  
\section{Conclusions}

In summary, by a combined experimental and \textit{ab-initio} theoretical study, the microstructure and deformability of the Al-3wt\%Nd alloy and the inherent mechanical properties, electron structures and chemical bonds of $\alpha$-$\AleNdt$ are studied comprehensively. The Al-3wt\%Nd alloy is composed of the pre-eutectic $\alpha$-Al matrix with a volume fraction of $\sim$92\% and the eutectic microstructure constituted by the alternately arranged $\alpha$-Al and $\alpha$-$\AleNdt$ phases. Under the asymmetric cold rolling, the Al-3wt\%Nd alloy possesses excellent plastic deformability. With the plastic deformation, the $\alpha$-Al matrix is elongated significantly along the rolling direction, and the eutectic microstructure transforms from a cellular to a lamellar shape. Nevertheless, the morphology of $\alpha$-$\AleNdt$ is not changed obviously. Theoretical calculations show that $\alpha$-$\AleNdt$ exhibits excellent thermodynamic and elastic stabilities. By examining ideal tensile strength, elastic moduli, hardness and inherent brittleness-ductility, the hardness-brittleness of $\alpha$-$\AleNdt$ is quantitatively evaluated, accounting for the difficulty in plastic deformation and fragmentation during cold rolling. By a combined investigation of band structure, population analysis, topological analysis and crystal orbital Hamilton population, it is revealed that $\alpha$-$\AleNdt$ possesses two kinds of chemical bonds: the Nd-Al and Al-Al bonds. The former is a typical ionic bond with the electron transfer from Nd to Al, while the latter, originated from 3\textit{s}-3\textit{p} and 3\textit{p}-3\textit{p} interactions, is a weak covalent bond. The mixed chemical bond is responsible for the high hardness-brittleness of $\alpha$-$\AleNdt$. This work shows that the microstructure uniformity of the Al-3wt\%Nd alloy can be significantly improved by plastic deformation. Meanwhile, it is confirmed that the plastic deformation or fragmentation of $\alpha$-$\AleNdt$ is relatively difficult. As is known, the size of the eutectic $\alpha$-$\AleNdt$ phase can be tailored by tuning the processing parameters during the casting, \textit{e.g.}, supercooling degree. It, thus, should be essential for a multiple-process collaborative control, \textit{e.g.}, casting and plastic deformation, to fabricate the high-quality Al-Nd target. The study is expected to lay a foundation for the Al-3wt\%Nd alloy and thus catalyze the fabrication of high-quality Al-Nd targets and further the advanced large-size TFT-LCD panels.
 
\section{Acknowledgments}
This work is supported by the National Key R\&D Program of China (2022YFB3504401).

\appendix
\section{Calculations of isotropic elastic moduli}\label{appen:1}
\setcounter{equation}{0}
\renewcommand{\theequation}{A\arabic{equation}}
Isotropic bulk modulus ($B$), shear modulus ($G$), Young’s modulus ($E$) and Poisson’s ratio ($\nu$) are calculated by Voigt–Reuss–Hill (V-R-H) approximation. For the orthorhombic system, the Voigt bulk modulus ($B_{\mathrm{V}}$) and shear modulus ($G_{\mathrm{V}}$) can be calculated by~\cite{RN466}:
\begin{equation}
B_{\mathrm{V}} = \frac{1}{9} \left[2(C_{12}+C_{13}+C_{23}) + C_{11}+C_{22}+C_{33}\right]
\end{equation}
\begin{align}
G_{\mathrm{V}} =& \frac{1}{15} \left[ C_{11} + C_{22}+C_{33} - C_{12} - C_{13} - C_{23})\right]  \nonumber \\
                & + \frac{1}{5} \left[(C_{44}+C_{55}+C_{66})\right]
\end{align}
The Reuss bounds of bulk modulus ($B_{\mathrm{R}}$) and shear modulus ($G_{\mathrm{R}}$) are as follows :
\begin{align}
B_{R} = & \Delta[C_{11}(C_{22} + C_{33}-2C_{23}) + C_{22}(C_{33}-2C_{13}) \nonumber \\
       &-2C_{33}C_{12}+C_{12}(2C_{23}-C_{12}) + C_{13}(2C_{12}-C_{13}) \nonumber \\
       &+C_{23}(2C_{13}-C_{23})]^{-1}
\end{align}
\begin{align}
G_{R}=& 15\{4[C_{11}(C_{22} + C_{33}+C_{23})+C_{22}(C_{33} + C_{13}) + C_{33}C_{12}  \nonumber \\
      &-C_{12}(C_{23}+C_{12})-C_{13}(C_{12}+C_{13})-C_{23}(C_{13}+C_{23})]/\Delta  \nonumber \\
      &+3[(1/C_{44})+(1/C_{55})+(1/C_{66})]\}^{-1}
\end{align}
where
\begin{align}
\Delta&=C_{13}(C_{12}C_{23}-C_{13}C_{22})+C_{23}(C_{12}C_{13}-C_{23}C_{11})  \nonumber \\
&+C_{33}(C_{11}C_{22}-C_{12}^{2}).
\end{align}

For the cubic system, the formula of $B_{\mathrm{V}}$ and $G_{\mathrm{V}}$ are as follows~\cite{RN466}:
\begin{equation}
G_{\mathrm{V}} = \frac{1}{5} [\left(C_{11}-C_{12})+3C_{44}\right]
\end{equation}
\begin{equation}
B_{\mathrm{V}} = \frac{1}{3} \left(C_{11}+2C_{12}\right)
\end{equation}
and the formula of $B_{\mathrm{R}}$ and $G_{\mathrm{R}}$ are as follows :
\begin{equation}
G_{\mathrm{R}} = \frac{5C_{44}(C_{11}-C_{12})} {3(C_{11}-C_{12})+4C_{44}}
\end{equation}
\begin{equation}
B_{\mathrm{R}} = \frac{1}{3} \left(C_{11}+2C_{12}\right)
\end{equation}
For the hexagonal system~\cite{RN466}, $B_{\mathrm{V}}$ and $G_{\mathrm{V}}$ are as follows:
\begin{equation}
B_{\mathrm{V}} = \frac{2}{9} \left(C_{11}+C_{12}+\frac{C_{33}}{2}+2C_{13}\right)
\end{equation}
\begin{equation}
G_{\mathrm{V}} = \frac{1}{30} (7C_{11}-5C_{12} + 12C_{44} + 2C_{33} - 4C_{13})
\end{equation}
and the formula of $B_{\mathrm{R}}$ and $G_{\mathrm{R}}$ are as follows :
\begin{equation}
B_{\mathrm{R}} = \frac{(C_{11}+C_{12})C_{33}-2C_{13}^{2}} {C_{11}+C_{12}+2C_{33}-4C_{13}}
\end{equation}
{\small
\begin{align}
G_{R}=\frac{5}{2}\left\{\frac{[(C_{11}+C_{12})C_{33}-2C_{13}^{2}]C_{44}C_{66}}{3B_{V}C_{44}C_{66}+[(C_{11}+C_{12})C_{33}-2C_{13}^{2}](C_{44}+C_{66})}\right\} 
\end{align}  
}

In the Voigt–Reuss–Hill (V-R-H) approximation, the average values of $B$ and $G$ can be calculated by:
\begin{equation}
B={\frac{1}{2}} (B_{\mathrm{V}} + B_{\mathrm{R}})
\end{equation}
\begin{equation}
G={\frac{1}{2}} (G_{\mathrm{V}}+G_{\mathrm{R}})
\end{equation}
and Young's modulus ($E$) and Poisson's ratio (\textit{$\nu$}) can be obtained by:
\begin{equation}
E={\frac{9GB}{G+3B}}
\end{equation}
\begin{equation}
\nu = {\frac{3B - 2G}{2(3B + G)}}
\end{equation}

\textbf{Reference}
\bibliographystyle{apsrev4-1}

\bibliography{ref}
\end{document}


\title{\textbf{Supplementary Materials to} \\``Deformability, inherent brittleness-ductility and chemical bonding of $\AleNdt$ in Al-Nd sputtering target material"}

\author{Xue-Qian Wang}
\affiliation{%
Key Laboratory for Anisotropy and Texture of Materials (Ministry of Education), School of Material Science and Engineering, Northeastern University, Shenyang 110819, China.
}%
\author{Run-Xin Song}
\affiliation{%
Key Laboratory for Anisotropy and Texture of Materials (Ministry of Education), School of Material Science and Engineering, Northeastern University, Shenyang 110819, China.
}%

\author{Xu Guan}
\affiliation{%
Key Laboratory for Anisotropy and Texture of Materials (Ministry of Education), School of Material Science and Engineering, Northeastern University, Shenyang 110819, China.
}%

 \author{Shuan Li}
 \affiliation{%
 National Engineering Research Center for Rare Earth, GRIREM Advanced Materials Co., Ltd., Beijing 100088, China
 }%

\author{Shuchen Sun}
\affiliation{%
School of Metallurgy, Northeastern University, Shenyang 110819, China.
}%

 \author{Hongbo Yang}
 \affiliation{%
 National Engineering Research Center for Rare Earth, GRIREM Advanced Materials Co., Ltd., Beijing 100088, China
 }%

 \author{Daogao Wu}
 \email{wudaogao@grirem.com (D.G. Wu)}
 \affiliation{%
 National Engineering Research Center for Rare Earth, GRIREM Advanced Materials Co., Ltd., Beijing 100088, China
 }%

\author{Ganfeng Tu}
\affiliation{%
School of Metallurgy, Northeastern University, Shenyang 110819, China.
}%

\author{Song Li}
\email{lis@atm.neu.edu.cn (S. Li)}
\affiliation{%
Key Laboratory for Anisotropy and Texture of Materials (Ministry of Education), School of Material Science and Engineering, Northeastern University, Shenyang 110819, China.
}%

\author{Hai-Le Yan}
\email{yanhaile@mail.neu.edu.cn (H.-L. Yan)}
\affiliation{%
Key Laboratory for Anisotropy and Texture of Materials (Ministry of Education), School of Material Science and Engineering, Northeastern University, Shenyang 110819, China.
}%

\author{Liang Zuo}
\affiliation{%
Key Laboratory for Anisotropy and Texture of Materials (Ministry of Education), School of Material Science and Engineering, Northeastern University, Shenyang 110819, China.
}%

\maketitle	
\hspace*{-1.25em}
\textbf{Supplementary Caption:}\\
The Supplementary Materials contain the detailed crystal structure information of $\alpha$-Al$_{11}$Nd$_{3}$, the relaxed structure file in the format of POSCAR by ab-initio calculation and the zoomed eutectic microstructure of the casted Al-3wt\% alloy.
\newline

\hspace*{-1.25em}
\textbf{Content:}\\
Section I: Crystal structure information of $\alpha$-Al$_{11}$Nd$_{3}$.\\
Section II: Relaxed structural model of $\alpha$-Al$_{11}$Nd$_{3}$.\\
Section III: Eutectic microstructure of the casted Al-3wt\% alloy.

\newpage

\section{Crystal structure information of $\alpha$-Al$_{11}$Nd$_{3}$}

\begin{table}[hbtp]
    \centering
    \caption{\textbf{Crystal structure information of Al$_{11}$Nd$_{3}$.} The space group is \textit{Immm} (group number: 71).}
    \begin{tabular}{cccccccc}
    \hline
         Atom & Label &  Wykoff & x & y &  z &  Occupancy\\
         \hline
         Nd & \Nd{I} & \textit{2a} & 0 & 0 & 0 & 1  & \\
         Nd & \Nd{II} & \textit{4i} & 0 & 0 & 0.81738 &  1 & \\
         Al & \Al{I} & \textit{2d} & 1/2 & 0 & 1/2 & 1  & \\
         Al & \Al{II} & \textit{4h} & 0 & 0.78388 & 1/2 & 1  & \\
         Al & \Al{III} & \textit{8l} & 0 & 0.63047 & 0.33407 & 1  & \\
         Al & \Al{IV} & \textit{8l} & 0 & 0.72599 & 0.86271 & 1  & \\ 
         \hline
    \end{tabular}

    \label{tab:my_label}
\end{table}

\section{Relaxed structural model of $\alpha$-Al$_{11}$Nd$_{3}$}

\noindent
Al11Nd3 in POSCAR format\\
 1.000000 \\ 
    4.3801096498628382    0.0000000000000000    0.0000000000000000 \\
    0.0000000000000000   10.0260238680764271    0.0000000000000000 \\
    0.0000000000000000    0.0000000000000000   13.0315663977200114 \\
   Nd   Al \\
     6    22 \\
Direct \\
    0.0000000000000000    0.0000000000000000    0.0000000000000000     Nd1 \\
    0.5000000000000000    0.5000000000000000    0.8176408510047521     Nd2 \\
    0.5000000000000000    0.5000000000000000    0.1823591489952479     Nd3 \\
    0.5000000000000000    0.5000000000000000    0.5000000000000000     Nd4 \\
    0.0000000000000000    0.0000000000000000    0.3176408510047521     Nd5 \\
    0.0000000000000000    0.0000000000000000    0.6823591489952479     Nd6 \\
    0.0000000000000000    0.5000000000000000    0.0000000000000000     Al1 \\
    0.5000000000000000    0.2852782166261220    0.0000000000000000     Al2 \\
    0.5000000000000000    0.7147217833738780    0.0000000000000000     Al3 \\
    0.0000000000000000    0.6312487585391187    0.3338148796572754     Al4 \\
    0.0000000000000000    0.3687512414608813    0.6661851203427246     Al5 \\
    0.0000000000000000    0.3687512414608813    0.3338148796572754     Al6 \\
    0.0000000000000000    0.6312487585391187    0.6661851203427246     Al7 \\
    0.0000000000000000    0.7254842833147103    0.1365571627913562     Al8 \\
    0.0000000000000000    0.2745157166852897    0.8634428372086438     Al9 \\
    0.0000000000000000    0.2745157166852897    0.1365571627913562    Al10 \\
    0.0000000000000000    0.7254842833147103    0.8634428372086438    Al11 \\
    0.5000000000000000    0.0000000000000000    0.5000000000000000    Al12 \\
    0.0000000000000000    0.7852782166261220    0.5000000000000000    Al13 \\
    0.0000000000000000    0.2147217833738782    0.5000000000000000    Al14 \\
    0.5000000000000000    0.1312487585391187    0.8338148796572754    Al15 \\
    0.5000000000000000    0.8687512414608813    0.1661851203427247    Al16 \\
    0.5000000000000000    0.8687512414608813    0.8338148796572754    Al17 \\
    0.5000000000000000    0.1312487585391187    0.1661851203427247    Al18 \\
    0.5000000000000000    0.2254842833147102    0.6365571627913562    Al19 \\
    0.5000000000000000    0.7745157166852897    0.3634428372086438    Al20 \\
    0.5000000000000000    0.7745157166852897    0.6365571627913562    Al21 \\
    0.5000000000000000    0.2254842833147102    0.3634428372086438    Al22 \\

\section{Microstructure of the casted Al-2at.\% alloy.}

\begin{figure}[htbp]
\centering
    \includegraphics[width=0.5\linewidth]{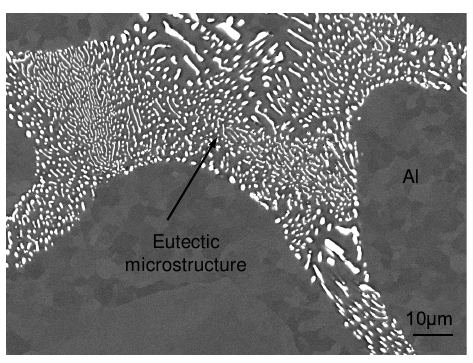}
    \caption{\textbf{Backscattered electron (BSE) image of the casted Al-3wt\%Nd alloy.} The compounds with the white color is the eutectic $\alpha$-Al$_{11}$Nd$_{3}$ phase.}
    \label{fig:coldrolling}
\end{figure}